\documentclass[aps,prb,showpacs,superscriptaddress]{revtex4}
\usepackage{times}
\usepackage{epsfig}
\usepackage{graphicx}
\usepackage{amsmath}
\begin{document}
\title{Parametric spin excitations in lateral quantum dots.}
\author{Jamie D. Walls}
\email{jwalls@fas.harvard.edu}
\affiliation{Department of Chemistry
and Chemical Biology,
 Harvard University, Cambridge, MA 02138}
\date{\today}
\begin{abstract}
In this work, the spin dynamics of a single electron under
parametric modulation of a lateral quantum dot's electrostatic
potential in the presence of spin-orbit coupling is investigated.
Numerical and theoretical calculations demonstrate that, by
squeezing and/or moving the electron's wave function, spin rotations
with Rabi frequencies on the order of tens of megahertz can be
achieved with experimentally accessible parameters in both parabolic
and square lateral quantum dots.  Applications of parametric
excitations for determining spin-orbit coupling parameters and for
increasing the spin polarization in the electronic ground are
demonstrated.
\end{abstract}
%\date{07/17/2000}
\maketitle

\section{Introduction}
Single spin operations are an important component for possible
realizations of quantum dot-based quantum computers\cite{Loss98}. In
principle, single spin manipulations can be performed using electron
spin resonance (ESR) techniques with the application of oscillating
magnetic fields.  A recent experiment\cite{Koppens06} utilizing ESR
methods was used to generate spin rotations in a single electron
lateral quantum dot at low static magnetic field strengths, where
the required frequencies were on the order of hundreds of megahertz.
However, extending ESR methods to microwave frequencies at low
temperatures while minimizing the accompanying AC electric field in
the semiconductor quantum dots is experimentally challenging.

Instead of considering the AC electric fields as something to be
avoided, AC electric fields can be used to control the electron spin
in a quantum dot.  Modulations of the electrostatic potential can
readily couple to the electronic degree of freedom in a
semiconductor quantum dot and can be performed on the nanosecond
timescale\cite{Petta04,Petta05}, which is roughly the timescale
associated with the Zeeman energy in static fields on the order of a
Tesla for GaAs quantum dots. However, in order to translate
electronic control into spin control, there must exist an additional
coupling between the spin and electronic degrees of freedom.

 The two main methods of coupling
the spin and electronic degrees of freedom are either through the
Zeeman interaction or through spin-orbit coupling. Consider first
the Zeeman interaction, which is given by
$\widehat{H}_{Z}=\vec{B}(\vec{r})\cdot\left(\widehat{g}(\vec{r})\mu_{B}\vec{S}\right)$,
where $\vec{B}(\vec{r})$ is the applied magnetic field,
$\widehat{g}(\vec{r})$ is the effective g-tensor, $\mu_{B}$ is the
Bohr magneton, and $\widehat{S}$ is the electron's spin vector.
Previous experimental work has demonstrated that electric fields can
be used to move the electron around in the presence of a spatially
dependent g-tensor, $\widehat{g}(\vec{r})$, in order to generate an
effective time-dependent Zeeman interaction,
$\widehat{H}_{Z}(t)=\hbar\vec{\omega}(t)\cdot\vec{S}$, which was
used to perform spin rotations\cite{Salis01,Kato03}.
 Alternatively, $\widehat{H}_{Z}(t)$ can also be generated by using
an electric field to move an electron around in the presence of a
spatially inhomogeneous magnetic field\cite{Tokura06}.

  Besides
manipulating the effective Zeeman interaction, modulating the
electrostatic potential can be used to manipulate the electron's
spin through the spin-orbit interaction. Previous theoretical
studies have proposed using electric
 fields to generate spin rotations via an electric dipole spin resonance (EDSR);  theoretical calculations indicate that EDSR methods can be used
 to rotate an electron's spin on the order of tens of nanoseconds
 for electrons in
 quantum wells\cite{Rashba03,Rashba03a,Efros06} and quantum dots\cite{Golovach06,Flindt06}.  For parabolic quantum dots, application of an electric field is
equivalent to modulating the center of the quantum dot which, in the
presence of
 spin-orbit coupling, is able to induce spin transitions when the modulation frequency is equal to the Zeeman frequency.  Besides utilizing
 EDSR for spin excitation, electric fields can also be used to
 modulate the Rashba spin-orbit coupling constant\cite{Nitta97}
 in order to perform combined spin and orbital excitations
 in quantum dots\cite{Debald05}.

In this work, the spin dynamics under the combined action of
spin-orbit coupling and parametric modulations of the electrostatic
confining potential (such as squeezing and moving the electronic
wave function) in both parabolic and square quantum dots is
examined. Theoretical and numerical calculations utilizing Floquet
theory and effective Hamiltonian theory are used to demonstrate that
parametric modulations of the electrostatic potential can generate
single spin rotations on the timescale of tens of nanoseconds using
experimentally accessible parameters. In order to maximize the
amplitude of the Rabi oscillations, the modulation frequency must be
chosen with precision on the order of the Rabi frequency (order of
1-10 MHz).  Such parametric excitations therefore can provide better
energy resolution of the
 quantum dot's energy levels over transport measurements, where the measured energy
  levels are often poorly resolved due to thermal effects. Besides performing
single spin rotations, we have shown that measurement of the
observed Rabi oscillations\cite{Engel01a,Engel02} and the required
modulation frequencies can be used  to determine the spin-orbit
coupling parameters in the quantum dot. Finally, we have
demonstrated that squeezing and expanding the electronic wave
function in a parabolic quantum dot can induce both orbital and spin
excitations, which can be used to increase the spin polarization of
the lowest electronic state.  It should be noted that relaxation is
neglected throughout most of the paper since the calculated Rabi
frequencies (on the order of 1-10 MHz) are at least one order of
magnitude greater than the experimentally observed $1/T_{1}$ values
found in GaAs quantum
dots\cite{Fujisawa02,Hanson03,Elzerman04,Amasha06}.

\section{Parametric Modulations and Floquet Theory}
Before studying the case of parametric excitations in lateral
quantum dots, the general formalism of using parametric modulations
to generate excitations in quantum systems is presented. Consider a
Hamiltonian which is a function of the parameter $\omega$:
\begin{eqnarray}
\widehat{H}(\omega)&=&\sum_{k}E_{k}(\omega)\frac{|k(\omega)\rangle\langle
k(\omega)|}{\langle
k(\omega)|k(\omega)\rangle}+\sum_{j<k}\frac{\lambda_{jk}(\omega)|j(\omega)\rangle\langle
k(\omega)|+\lambda_{kj}(\omega)|k(\omega)\rangle\langle
j(\omega)|}{\sqrt{\langle k(\omega)|k(\omega)\rangle\langle
j(\omega)|j(\omega)\rangle}}\nonumber\\
&=&\widehat{W}(\omega)\widetilde{H}_{0}(\omega)\widehat{W}^{\dagger}(\omega)
\label{eq:Href}
\end{eqnarray}
where
\begin{eqnarray}
\widetilde{H}_{0}(\omega)&=&\sum_{k}E_{k}(\omega)\frac{|k_{0}\rangle\langle
k_{0}|}{\langle
k_{0}|k_{0}\rangle}+\sum_{j<k}\frac{\lambda_{jk}(\omega)|j_{0}\rangle\langle
k_{0}|+\lambda_{kj}(\omega)|k_{0}\rangle\langle
j_{0}|}{\sqrt{\langle k_{0}|k_{0}\rangle\langle
j_{0}|j_{0}\rangle}}\\
\label{eq:hoo}
\widehat{W}(\omega)&=&\sum_{k}\frac{|k(\omega)\rangle\langle
k_{0}|}{\sqrt{\langle k(\omega)|k(\omega)\rangle\langle
k_{0}|k_{0}\rangle}} \label{eq:woo}
\end{eqnarray}
with $|k_{0}\rangle\equiv|k(\omega_0)\rangle$.  The transformation
$\widehat{W}(\omega)$ simply switches from the basis
$\{|k(\omega_{0})\rangle\}$ to the basis $\{|k(\omega)\rangle\}$.

In order to induce transitions between two arbitrary states
$|k(\omega)\rangle$ and $|j(\omega)\rangle$ efficiently, the matrix
element coupling these states must be modulated at a frequency
$\omega_{r}$ given by
$n\hbar\omega_{r}\approx|E_{j}(\omega)-E_{k}(\omega)|$ for integer
$n$. The time-dependent coupling between states can in principle be
generated by modulating the parameter $\omega$ of
$\widehat{H}(\omega)$. The propagator during modulation of $\omega$
can be written as:
\begin{eqnarray}
\widehat{U}(t)&=&T\exp\left(-\frac{\text{i}}{\hbar}\int^{t}_{0}\text{d}t'\widehat{H}(\omega(t'))\right)\nonumber\\
&=&T\exp\left(-\frac{\text{i}}{\hbar}\int^{t}_{0}\text{d}t'\widehat{W}(\omega(t'))\widetilde{H}_{0}(\omega(t'))\widehat{W}^{\dagger}(\omega(t'))\right)\nonumber\\
&=&\widehat{W}(\omega(t))T\exp\left(-\frac{\text{i}}{\hbar}\int^{t}_{0}\text{d}t'\left[-\text{i}\hbar\widehat{W}^{\dagger}(\omega(t'))\frac{\partial\widehat{W}(\omega(t'))}{\partial
t'}+\widetilde{H}_{0}(\omega(t'))\right]\right)\nonumber\\
&=&\widehat{W}(\omega(t))T\exp\left(-\frac{\text{i}}{\hbar}\int^{t}_{0}\text{d}t'\widetilde{H}(t')\right)
\label{eq:Hber}
\end{eqnarray}
where
\begin{eqnarray}
\label{eq:Hber1}
\widetilde{H}(t)&=&\widetilde{H}_{0}(\omega(t))-\text{i}\hbar\widehat{W}^{\dagger}(\omega(t))\frac{\partial\widehat{W}(\omega(t))}{\partial
t}\\
-\text{i}\hbar\widehat{W}^{\dagger}(\omega(t))\frac{\partial\widehat{W}(\omega(t))}{\partial
t}&=&\left(\sum_{k}V_{kk}(t)\frac{|k_{0}\rangle\langle
k_{0}|}{\langle
k_{0}|k_{0}\rangle}+\sum_{j<k}\frac{V_{jk}(t)|j_{0}\rangle\langle
k_{0}|+V_{kj}(t)|k_{0}\rangle\langle j_{0}|}{\sqrt{\langle
j_{0}|j_{0}\rangle\langle k_{0}|k_{0}\rangle}}\right)
\end{eqnarray}
with \begin{eqnarray} \text{for $j\neq k$}&&\nonumber\\
V_{kj}(t)&=&-\text{i}\hbar\frac{\partial{\omega(t)}}{\partial
t}\frac{\langle k(\omega(t))|\frac{\partial j(\omega(t))}{\partial
\omega(t)}\rangle}{\sqrt{\langle
j(\omega(t))|j(\omega(t))\rangle\langle
k(\omega(t))|k(\omega(t))\rangle}}\nonumber\\
&=&-\frac{\text{i}\hbar}{\Delta_{jk}(\omega(t))}\frac{\partial\omega(t)}{\partial
t}\frac{\langle
k(\omega(t))|\frac{\partial\widehat{H}(\omega(t))}{\partial
\omega(t)}|j(\omega(t))\rangle}{\sqrt{\langle
j(\omega(t))|j(\omega(t))\rangle\langle
k(\omega(t))|k(\omega(t))\rangle}}\\
\text{for $j=k$}&&\nonumber\\
V_{kk}(t)&=&-\text{i}\hbar\frac{\partial{\omega(t)}}{\partial
t}\frac{\langle k(\omega(t))|\frac{\partial k(\omega(t))}{\partial
\omega(t)}\rangle}{\langle k(\omega(t))|k(\omega(t))\rangle}
\end{eqnarray}
where $\Delta_{jk}(\omega(t))=E_{j}(\omega(t))-E_{k}(\omega(t))$.
 Note that $\int^{t}_{0}\text{d}t' V_{kk}(t')/\hbar$ is just the geometric
phase\cite{Sakurai} associated with the parametric modulation of
$\widehat{H}(\omega)$.

 In the following, we are
interested in the case where only small, periodic modulations to
$\omega$ are performed, i.e,
$\omega(t)=\omega+\delta\omega\sin(\omega_{r}t+\phi)$ where
$\delta\omega\ll \omega$, and $\omega_{r}$ and $\phi$ are the
modulation frequency and the initial phase respectively.   If the
energy difference between the states $|j_{0}\rangle$ and
$|k_{0}\rangle$ is much greater than the generated off-diagonal
matrix elements $(|\Delta_{jk}(\omega(0))|\gg |V_{jk}|)$, then the
two states will remain uncoupled unless $|\Delta_{jk}(\omega(0))|$
is equal to a multiple of $\hbar\omega_{r}$, i.e.,
$n\hbar\omega_{r}\approx |\Delta_{jk}(\omega(0))|$ where $n$ is an
integer.  In order to gain insight into the dynamics under periodic
modulation of the Hamiltonian, Floquet theory\cite{Shirley65} can be
used to solve for $\widehat{U}(t)$ in Eq. (\ref{eq:Hber}). Rewriting
$\widetilde{H}(t)$ in Eq. (\ref{eq:Hber1}) as
$\widetilde{H}(t)=\widetilde{H}_{0}(\omega(0))+\widehat{V}(t)$, with
$\widetilde{H}_{0}(\omega(0))$ given by Eq. (\ref{eq:hoo}) and
$\widehat{V}(t)=\sum_{m}\widehat{V}_{m}\exp(\text{i}m\omega_{r}t)$,
the effective propagator can be written as:
\begin{eqnarray}
\widehat{W}^{\dagger}(t)\widehat{U}(t)&=&T\exp\left(-\frac{\text{i}}{\hbar}\int^{t}_{0}\text{d}t'\widetilde{H}(t')\right)\\
&=&T\exp\left(-\frac{\text{i}}{\hbar}\int^{t}_{0}\text{d}t'\left[\widetilde{H}_{0}(\omega(0))+\sum_{m}\widehat{V}_{m}\exp(\text{i}m\omega_{r}t)\right]\right)\nonumber\\
&=&\sum_{n=-\infty}^{\infty}\exp(\text{i}n\omega_{r}t)\widehat{U}_{n}(t)
\label{eq:fourierdeco}
\end{eqnarray}
where Eq. (\ref{eq:fourierdeco}) represents the Fourier
decomposition of the propagator $\widehat{U}(t)$ into the Fourier
operators, $\widehat{U}_{n}(t)$, which satisfy:
\begin{eqnarray}
\frac{\text{d}\widehat{U}_{n}(t)}{\text{d}t}&=&-\frac{\text{i}}{\hbar}\left(n\omega_{r}\widehat{1}+\widetilde{H}_{0}(\omega(0))\right)\widehat{U}_{n}(t)-\frac{\text{i}}{\hbar}\sum_{m}\widehat{V}_{m}\widehat{U}_{n-m}(t)
\label{eq:fourier}
\end{eqnarray}
with the initial conditions chosen to be
$\widehat{U}_{0}(0)=\widehat{1}$ and $\widehat{U}_{n\neq
0}=\widehat{0}$ in order that $\widehat{U}(0)=\widehat{1}$.  The
solution to Eq. (\ref{eq:fourier}) is given by
${\bf{\widehat{U}}}(t)=\exp(-\text{i}\widehat{H}_{F}t/\hbar){\bf{\widehat{U}}}(0)$,
where ${\bf{\widehat{U}}}(t)$ can be formally represented as a
vector in Floquet space,
${\bf{\widehat{U}}}(t)=\sum_{n=-\infty}^{\infty}\widehat{U}_{n}(t)|n_{F}\rangle$
with ${\bf{\widehat{U}}}(0)=\widehat{1}|0_{F}\rangle$ [where the
various Floquet states $|n_{F}\rangle$ satisfy $\langle
n_{F}|m_{F}\rangle=\delta_{n_{F},m_{F}}$], and $U_{n}(t)=\langle
n_{F}|\exp\left(-\text{i}\widehat{H}_{F}t/\hbar\right)|0_{F}\rangle$.
The time-independent Floquet Hamiltonian, $\widehat{H}_{F}$, is
defined by:
\begin{eqnarray}
\langle n_{F}|\widehat{H}_{F}|m_{F}\rangle&=&\langle
n_{F}|\widetilde{H}_{0,F}+\widehat{N}_{F}\hbar\omega_{r}|m_{F}\rangle+\langle
n_{F}|\widehat{V}_{F}|m_{F}\rangle\nonumber\\
&=&
\delta_{n_{F},m_{F}}\left(\widetilde{H}_{0}(\omega(0))+n_{F}\hbar\omega_{r}\widehat{1}\right)+\sum_{k}\delta_{n_{F}-k,m_{F}}\widehat{V}_{k}
\end{eqnarray}
where the states in Floquet space are denoted by $|j,n_{F}\rangle$,
with
$(\widetilde{H}_{0,F}+\widehat{N}_{F}\hbar\omega_{r})|j,n_{F}\rangle=(E_{j}(\omega(0))+n_{F}\hbar\omega_{r})|j,n_{F}\rangle$.

By going into Floquet space, the initial time-dependent Hamiltonian,
$\widetilde{H}(t)$, has been replaced by the time-independent
Floquet Hamiltonian, $\widehat{H}_{F}$; however, in order to
calculate ${\bf{\widehat{U}}}(t)$, we must now exponentiate
$\widehat{H}_{F}$, which is an infinite dimensional matrix in
Floquet space, since $\langle n_{F}|\widehat{H}_{F}|m_{F}\rangle$ is
defined for all $n_{F},m_{F}\in\{-\infty,\infty\}$. However, the
propagator can be simplified if there exists a finite dimensional
subspace of nearly degenerate states, $Q$, which is weakly coupled
to states outside of $Q$.  In this case, the dynamics within $Q$ can
be separated from the rest of the Floquet space by constructing an
effective Hamiltonian\cite{Shavitt80,Tannoudji} within the subspace
$Q$ by treating the coupling to states outside of $Q$
perturbatively. For example, take
$Q=\{|j,n_{F}\rangle,|k,(n+p)_{F}\rangle\}$ with
$E_{j}(\omega(0))\approx E_{k}(\omega(0))+p\hbar\omega_{r}$, and
assume that $Q$ is weakly coupled to states outside of $Q$ (i.e.,
$|\langle j,n_{F}|\widehat{V}_{F}|q,m_{F}\rangle|\ll
|E_{j}(\omega(0))-E_{q}(\omega(0))+(n-m)_{F}\hbar\omega_{r}|$ and
$|\langle k,(n+p)_{F}|\widehat{V}_{F}|q,m_{F}\rangle|\ll
|E_{k}(\omega(0))-E_{q}(\omega(0))+(n+p-m)_{F}\hbar\omega_{r}|$), an
effective Hamiltonian can be written in the subspace $Q$ by using a
unitary transformation, $\exp\left(\widehat{S}_{F}\right)$, where
$\exp\left(\widehat{S}_{F}\right)\widehat{H}_{F}\exp\left(-\widehat{S}_{F}\right)=\widehat{H}_{F}^{\text{EFF}}$;
the state $|j,n_{F}\rangle$ is only coupled to the state
$|k,(n+p)_{F}\rangle$ in $\widehat{H}_{F}^{\text{EFF}}$, i.e.,
$\langle p,r_{F}|\widehat{H}_{F}^{\text{EFF}}|m,q_{F}\rangle\neq 0$
only if $|m,q_{F}\rangle,|p,r_{F}\rangle\in Q$. A perturbation
series for $\widehat{S}_{F}$ in powers of $\widehat{V}_{F}$,
$\widehat{S}_{F}=\sum_{n=1}^{\infty}\widehat{S}_{F}^{(n)}$, can be
constructed and is given in Appendix A.

Using $\widehat{H}^{\text{EFF}}_{F}$, the propagator,
$\widehat{U}(t)$ can finally be written as:
\begin{eqnarray}
\widehat{W}^{\dagger}(t)\widehat{U}(t)&=&\sum_{n=\infty}^{\infty}\exp(\text{i}\omega_{r}t)U_{n}(t)\nonumber\\
&=&\sum_{n=-\infty}^{\infty}\exp(\text{i}n\omega_{r}t)\langle
n_{F}|\exp\left(-\widehat{S}_{F}\right)\exp\left(-\frac{\text{i}}{\hbar}\widehat{H}_{F}^{\text{EFF}}t\right)\exp\left(\widehat{S}_{F}\right)|0_{F}\rangle\nonumber\\
&\approx&\sum_{n=-\infty}^{\infty}\exp(\text{i}n\omega_{r}t)\left(\langle
n_{F}|\exp\left(-\frac{\text{i}}{\hbar}\widehat{H}_{F}^{\text{EFF}}t\right)|0_{F}\rangle-\langle
n_{F}|[\widehat{S}^{(1)}_{F},\exp\left(-\frac{\text{i}}{\hbar}\widehat{H}_{F}^{\text{EFF}}t\right)]|0_{F}\rangle\right)\nonumber\\
\end{eqnarray}

The resulting propagator, projected onto the $\{|j_{0}\rangle,
|k_{0}\rangle\}$ subspace (where
$\widehat{P}_{s}=|j_{0}\rangle\langle j_{0}|+|k_{0}\rangle\langle
k_{0}|$), is given by:
\begin{eqnarray}
\widehat{P}_{s}\widehat{W}^{\dagger}(t)\widehat{U}(t)\widehat{P}_{s}&=&\widehat{Z}(\omega_{r}t)\left(\widehat{U}^{jk}(t)-\sum_{n\neq
0}\exp(\text{i}n\omega_{r}t)\widehat{X}_{n}\widehat{U}^{jk}(t)-\widehat{U}^{jk}(t)\widehat{X}_{n}\right)
\end{eqnarray}
where
\begin{eqnarray}
\widehat{X}_{n}&=&\left(\begin{array}{cc}\frac{\langle
j,n_{F}|\widehat{V}_{F}|j,0_{F}\rangle}{n\omega_{r}}&\frac{\langle
j,(n-p)_{F}|\widehat{V}_{F}|k,0\rangle}{\Delta_{jk}+(n-p)\hbar\omega_{r}}\\
\frac{\langle
k,(n+p)_{F}|\widehat{V}_{F}|j,0\rangle}{(n+p)\hbar\omega_{r}-\Delta_{jk}}&\frac{\langle k,n_{F}|\widehat{V}_{F}|k,0_{F}\rangle}{n\omega_{r}}\end{array}\right)\nonumber\\\nonumber\\
\widehat{Z}(\omega_{r}t)&=&\left(\begin{array}{cc}1&0\\0&\exp(\text{i}\omega_{r}t)\end{array}\right)
\end{eqnarray}
and
$\widehat{U}^{jk}(t)=\exp\left(-\text{i}\widehat{H}^{jk}_{\text{EFF}}t/\hbar\right)$,
where $\widehat{H}^{jk}_{\text{EFF}}$ is the effective Hamiltonian
in the $\{|j,0_{F}\rangle,|k,p_{F}\rangle\}$ Floquet subspace, i.e.,
$\widehat{H}^{jk}_{\text{EFF}}=\left(|j,0_{F}\rangle\langle
j,0_{F}|+|k,p_{F}\rangle\langle
k,p_{F}|\right)\widehat{H}^{\text{EFF}}_{F}\left(|j,0_{F}\rangle\langle
j,0_{F}|+|k,p_{F}\rangle\langle k,p_{F}|\right)$.\label{sec:goo0}
\section{Parametric modulations in lateral, parabolic quantum dots}
\label{sec:goo1} Using the formalism presented in Section
\ref{sec:goo0}, we are now ready to begin studying parametric
excitation in a single-electron, lateral quantum dot, which is taken
to lie in the XY plane with the electron's wave function strongly
confined along the $\widehat{z}$ direction.  In the following, only
an in-plane magnetic field,
$\vec{B}=B\cos(\theta)\widehat{x}+B\sin(\theta)\widehat{y}$, will be
considered, which allows one to neglect orbital effects associated
with an out of plane magnetic field.  The Hamiltonian for a
single-electron lateral quantum dot defined by the electrostatic
potential $\widehat{V}(\widehat{X},\widehat{Y})$, in the presence of
spin-orbit coupling (for the moment, only Rashba\cite{Bychkov84} and
linear Dresselhaus\cite{Dresselhaus55} are considered), is given by:
\begin{eqnarray}
\label{eq:Hintro}
\widehat{H}&=&\widehat{H}_{0}+\widehat{H}_{SO}\\
\widehat{H}_{0}&=&\frac{\widehat{P}_{X}^{2}}{2m^{*}}+\frac{\widehat{P}_{Y}^{2}}{2m^{*}}+\widehat{V}(\widehat{X},\widehat{Y})-\frac{\hbar\omega_{Z}}{2}\widehat{\sigma}_{Z}\\
\label{eq:Hintro1}
\widehat{H}_{SO}&=&\frac{\widehat{P}_{X}}{\hbar}\left(\zeta_{1}(-\theta)\widehat{\sigma}_{X}+\zeta_{2}(\theta)\widehat{\sigma}_{Z}\right)
-\frac{\widehat{P}_{Y}}{\hbar}\left(\zeta_{2}(-\theta)\widehat{\sigma}_{X}+\zeta_{1}(\theta)\widehat{\sigma}_{Z}\right)
\label{eq:Hintro2}
\end{eqnarray}
where $\hbar\omega_{Z}=|g\mu_{B}\vec{B}|$,
$\zeta_{1}(\theta)=\alpha\cos(\theta)-\beta\sin(\theta)$,
$\zeta_{2}(\theta)=\alpha\sin(\theta)-\beta\cos(\theta)$ [where
$\alpha$ and $\beta$ are the Rashba and linear Dresselhaus coupling
constants respectively], and the vector potential,
$\vec{A}=\widehat{Z}\sin(\theta)\widehat{x}-\widehat{Z}\cos(\theta)\widehat{y}$,
was chosen. Due to the strong confinement along the
$\widehat{z}$-direction, all terms linear in $\widehat{Z}$ have been
truncated/removed from $\widehat{H}$ in Eqs.
(\ref{eq:Hintro})-(\ref{eq:Hintro2}), with terms quadratic in
$\widehat{Z}$ being incorporated into the the confining potential
along the $\widehat{z}$-direction.  The electrostatic potential,
$\widehat{V}(\widehat{X},\widehat{Y})$, mostly results from voltages
applied to surface gates above the 2DEG, which confines the electron
within the quantum dot;  changing the voltages of the surface gates
can change $\widehat{V}(\widehat{X},\widehat{Y})$. The spin
quantization axis has been taken to be along the direction of the
in-plane magnetic field. The eigenstates of $\widehat{H}_{0}$ are
denoted by $|n,\pm\rangle$, where
$\widehat{H}_{0}|n,\pm\rangle=(E_{n}\mp\hbar\omega_{Z}/2)|n,\pm\rangle$.
In the presence of spin-orbit coupling, the various $|n,\pm\rangle$
states are mixed;  however, if the confinement strength is much
larger than the spin-orbit coupling strength, i.e., $|\langle
n,\pm|\widehat{H}_{SO}|m,\pm\rangle|\ll |\Delta_{nm}|$ and $|\langle
n,\mp|\widehat{H}_{SO}|m,\pm\rangle|\ll | \Delta_{nm}\pm
\omega_{Z}|$ for all $n$ and $m$, $\widehat{H}_{SO}$ is suppressed
and can be treated as a perturbation to $\widehat{H}_{0}$.

 In this section, a lateral quantum dot defined by a parabolic electrostatic potential will be examined. Such parabolic potentials have enjoyed
tremendous success in describing transport and spectral properties
in lateral quantum dots\cite{Tarucha96,Maksym90,Merkt91}. The
electrostatic potential for a parabolic quantum dot can be written
as:
\begin{eqnarray}
\widehat{V}(\widehat{X},\widehat{Y})&=&\frac{m^{*}\omega^{2}_{X}}{2}\left(\widehat{X}-x_{c'}\right)^{2}+\frac{m^{*}\omega_{Y}^{2}}{2}\left(\widehat{Y}-y_{c'}\right)^{2}-eF_{X}\widehat{X}-eF_{Y}\widehat{Y}\nonumber\\
&=&\frac{m^*\omega^{2}_{X}}{2}\left(\widehat{X}-\frac{eF_{X}}{m^{*}\omega_{X}^{2}}-x_{c'}\right)^2+\frac{m^{*}\omega^{2}_{Y}}{2}\left(\widehat{Y}-\frac{eF_{Y}}{m^{*}\omega_{Y}^{2}}-y_{c'}\right)^{2}-\frac{e^{2}F_{X}^2}{2m^{*}\omega_{X}^2}-\frac{e^{2}F_{Y}^{2}}{2m^{*}\omega_{Y}^2}\nonumber\\
\label{eq:parpot}
\end{eqnarray}
where $\vec{r}_{c'}=x_{c'}\widehat{x}+y_{c'}\widehat{y}$ is the
center of the parabolic well, $\omega_{X}$ and $\omega_{Y}$ are the
oscillator frequencies of the parabolic dot, and $F_{X}$ and $F_{Y}$
are static electric fields which do not alter the energy levels of
the quantum dot but do shift the effective center of the quantum dot
to $\vec{r}_{c}=
[x_{c'}+eF_{X}/(m^{*}\omega^2_{X})]\widehat{x}+[y_{c'}+eF_{Y}/(m^{*}\omega^2_{Y})]\widehat{y}$.
For convenience, $\widehat{V}(\widehat{X},\widehat{Y})$ in Eq.
(\ref{eq:parpot}) was taken to be separable in the $\widehat{X}$ and
$\widehat{Y}$ degrees of freedom; that is, the principal axes of
$\widehat{V}(\widehat{X},\widehat{Y})$ were chosen for convenience
to be the same as those used in Eq. (\ref{eq:Hintro}) (if this is
not the case, inclusion of terms like
$\lambda\widehat{X}\widehat{Y}$ can be readily incorporated into the
following theory). The eigenstates of $\widehat{H}_{0}$ for
parabolic confinement,$|n,m,\pm\rangle$, satisfy
$\widehat{H}_{0}|n,m,\pm\rangle=\left[\hbar\omega_{X}(n+1/2)+\hbar\omega_{Y}(m+1/2)\mp\hbar\omega_{Z}/2\right]|n,m,\pm\rangle
$ and are centered about $\vec{r}_{c}$.  Under conditions of strong
confinement (where
$\hbar\omega_{X},\hbar\omega_{Y}\gg|\widehat{H}_{SO}|$), the states
$|0,0,+\rangle$ and $|0,0,-\rangle$ are approximately the two lowest
energy eigenstates.  Transitions between these two states correspond
to spin rotations within the ground electronic state.

\begin{figure}%[b!]
\includegraphics*[height=9.7cm,width = 12.7cm]{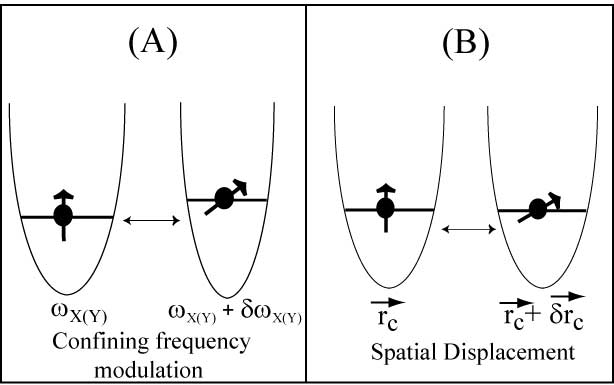}
\caption{Parametric modulations of a parabolic quantum dot's
confining potential by either modulating (A) the confining frequency
($\omega_{X}$ and/or $\omega_{Y}$) or (B) the center of the dot,
$\vec{r}_{c}=x_{c}\widehat{x}+y_{c}\widehat{y}$. In the presence of
spin-orbit coupling, if either $\delta\omega_{X(Y)}(t)$ or
$\delta\vec{r}_{c}(t)\equiv\delta x_{c}(t)\widehat{x}+\delta
y_{c}(t)\widehat{y}$ is modulated at roughly the energy difference
between $|0,0,+\rangle$ and $|0,0,-\rangle$, effective spin
rotations can be performed as depicted in Fig.
\ref{fig:fig1}.}\label{fig:fig1}
\end{figure}

As discussed in Section \ref{sec:goo0}, efficient transitions
between the states $|0,0,+\rangle$ and $|0,0,-\rangle$ in the
quantum dot can occur whenever the Hamiltonian is parametrically
modulated at roughly the energy difference between the two states.
There are two natural modulation parameters to choose from in
parabolic quantum dot potentials as shown in Fig. \ref{fig:fig1}:
either modulation of the oscillator frequencies
[$\omega_{X(Y)}\rightarrow\omega_{X(Y)}+\delta\omega_{X(Y)}$ in Fig.
\ref{fig:fig1}(A)], or modulation of the center of the quantum dot
[$\vec{r}_{c}\rightarrow\vec{r}_{c}+\delta\vec{r}_{c}$ in Fig.
\ref{fig:fig1}(B)] (note that modulation of $F_{X}$ and/or $F_{Y}$
is equivalent to modulating $\vec{r}_{c}$). Such parametric
modulations can in principle be performed by modulating the surface
gate voltages which define $\widehat{V}(\widehat{X},\widehat{Y})$.
Modulations of surface gate voltages have been experimentally
performed on the nanosecond timescale\cite{Petta04,Petta05}.
\subsection{Parametric Modulations of the Confining Strength in a
Parabolic Quantum Dot}

The modulation the the oscillator strengths of the quantum dot,
$\omega_{X(Y)}\rightarrow\omega_{X(Y)}(t)=\omega_{X(Y)}+\delta\omega_{X(Y)}(t)$,
corresponds to squeezing and expanding the electron's wave function
in a time-dependent manner, as illustrated in Fig.
\ref{fig:fig1}(A). Using Eq. (\ref{eq:Hber}), the effective
time-dependent Hamiltonian during modulation of $\omega_{X}$ and
$\omega_{Y}$ at a frequency $\omega_{r}$ is given by:
\begin{eqnarray}
\label{eq:Hparmod1}
\frac{\widetilde{H}(t)}{\hbar}&=&\omega_{X}(t)(a^{\dagger}_{X}a_{X}+1/2)+\omega_{Y}(t)(a^{\dagger}_{Y}a_{Y}+1/2)-\frac{\omega_{Z}}{2}\widehat{\sigma}_{Z}\nonumber\\
&+&\text{i}\sqrt{\frac{m^{*}\omega_{X}(t)}{2\hbar^3}}\left(a^{\dagger}_{X}-a_{X}\right)\left(\zeta_{1}(-\theta)\widehat{\sigma}_{X}+\zeta_{2}(\theta)\widehat{\sigma}_{Z}\right)+\frac{\text{i}}{4\omega_{X}(t)}\frac{\partial\omega_{X}(t)}{\partial
t}\left((a^{\dagger}_{X})^{2}-a_{X}^{2}\right)\nonumber\\
&-&\text{i}\sqrt{\frac{m^{*}\omega_{Y}(t)}{2\hbar^{3}}}\left(a^{\dagger}_{Y}-a_{Y}\right)\left(\zeta_{2}(-\theta)\widehat{\sigma}_{X}+\zeta_{1}(-\theta)\widehat{\sigma}_{Z}\right)+\frac{\text{i}}{4\omega_{Y}(t)}\frac{\partial\omega_{Y}(t)}{\partial
t}\left((a^{\dagger}_{Y})^{2}-a_{Y}^{2}\right)\nonumber\\
&+&\text{i}\frac{\partial\omega_{X}(t)}{\partial t}\frac{e
F_{X}}{\omega_{X}^{2}(t)}\sqrt{\frac{2}{\hbar
m^{*}\omega_{X}(t)}}\left(a_{X}^{\dagger}-a_{X}\right)+\text{i}\frac{\partial\omega_{Y}(t)}{\partial
t}\frac{eF_{Y}}{\omega_{Y}^{2}(t)}\sqrt{\frac{2}{\hbar m^{*}\omega_{Y}(t)}}\left(a_{Y}^{\dagger}-a_{Y}\right)\nonumber\\
&&\\
&\equiv&\frac{1}{\hbar}\left(\widetilde{H}_{0}(\omega_{X},\omega_{Y})+\sum_{m}\widehat{V}_{m}\exp(\text{i}m\omega_{r}t)\right)\label{eq:Hparmod}
\end{eqnarray}
where $a^{\dagger}_{X(Y)}$ and $a_{X(Y)}$ are the creation and
annihilation operators associated with the harmonic potential, and
$\omega_{X(Y)}(t)=\omega_{X(Y)}+\delta\omega_{X(Y)}\sin(\omega_{r}t+\phi_{X(Y)})$
is the time-dependent oscillator frequency.  Note that
$\widetilde{H}_{0}(\omega_{X},\omega_{Y})$ contains the
time-independent contributions to the energy from the harmonic
potential and the Zeeman and spin-orbit interactions.  In the
presence of a static electric field (i.e., $F_{X(Y)}\neq 0$),
modulations of $\omega_{X}$ and $\omega_{Y}$ also results in a
modulation of $\vec{r}_{c}$, which leads to the terms linear in
$a_{X(Y)}$ and $a^{\dagger}_{X(Y)}$ in Eq. (\ref{eq:Hparmod1}).

Efficient transitions between the Floquet states
$|1\rangle\equiv|0,0,+,1_{F}\rangle$ and
$|2\rangle\equiv|0,0,-,0_F\rangle$ can be performed if the
modulation frequency, $\omega_{r}$, equals the energy difference
between states $|1\rangle$ and $|2\rangle$, i.e.,
$\omega_{r}\approx\omega_{Z}$. Using Eq. (\ref{eq:Hber}), the
effective Hamiltonian, $\widehat{H}^{12}_{\text{EFF}}$, in the
presence of parametric modulation of the oscillator strengths at
$\omega_{r}\approx \omega_{Z}$ is given [up to
$\delta\omega_{X(Y)}^{2}$]:
\begin{eqnarray}
\widehat{H}^{12}_{\text{EFF}}&=&\frac{\hbar}{2}\left(\omega_{r}-\omega_{Z}-\Delta^{12}_{Z}-\delta_{Z}\right)\widehat{\sigma}^{12}_{Z}+\delta_{+}\widehat{\sigma}^{12}_{+}+\delta_{-}\widehat{\sigma}^{12}_{-}
\label{eq:How}
\end{eqnarray}
where $\widehat{\sigma}^{12}_{Z}=|1\rangle\langle
1|-|2\rangle\langle 2|$, $\widehat{\sigma}^{12}_{+}=|1\rangle\langle
2|$, $\widehat{\sigma}^{12}_{-}=|2\rangle\langle 1|$, and
\begin{eqnarray}
\Delta^{12}_{Z}&=&\frac{m^{*}\omega_{Z}}{\hbar^{3}}\left(\frac{\omega_{X}\left(\zeta_{1}(-\theta)\right)^{2}}{\omega_{Z}^{2}-\omega^{2}_{X}}+\frac{\omega_{Y}\left(\zeta_{2}(-\theta)\right)^{2}}{\omega_{Z}^{2}-\omega_{Y}^{2}}\right)\nonumber\\
\delta_{Z}&=&\frac{m^{*}\omega_{Z}\delta\omega_{X}^2\left(\zeta_{1}(-\theta)\right)^{2}}{4\hbar^3\omega_{X}}\left(\frac{32\omega_{X}^8+4\omega_{X}^6\omega_{Z}^2-85\omega_{X}^4\omega_{Z}^4+27\omega_{X}^2\omega_{Z}^6+4\omega_{Z}^{8}}{(\omega_{X}^{2}-4\omega_{Z}^{2})(4\omega_{X}^2-\omega_{Z}^{2})^{2}(\omega_{X}^2-\omega_{Z}^{2})^{2}}\right)\nonumber\\
&+&\frac{m^{*}\omega_{Z}\delta\omega_{Y}^2\left(\zeta_{2}(-\theta)\right)^{2}}{4\hbar^3\omega_{Y}}\left(\frac{32\omega_{Y}^8+4\omega_{Y}^6\omega_{Z}^2-85\omega_{Y}^4\omega_{Z}^4+27\omega_{Y}^2\omega_{Z}^6+4\omega_{Z}^{8}}{(\omega_{Y}^{2}-4\omega_{Z}^{2})(4\omega_{Y}^2-\omega_{Z}^{2})^{2}(\omega_{Y}^2-\omega_{Z}^{2})^{2}}\right)\nonumber\\
&+&\frac{\omega_{Z}}{\hbar^4}\left((\delta
\omega_{X}eF_{X})^2\frac{(\omega_{X}^{2}+3\omega_{Z}^{2})(\zeta_{1}(-\theta))^2}{\omega_{X}^2\left(\omega_{Z}^{2}-\omega_{X}^2\right)^3}+(\delta
\omega_{Y}eF_{Y})^{2}\frac{(\omega_{Y}^{2}+3\omega_{Z}^{2})(\zeta_{2}(-\theta))^{2}}{\omega_{Y}^2\left(\omega_{Z}^{2}-\omega_{Y}^{2}\right)^3}\right)\nonumber\\
&+&2\frac{eF_{X}eF_{Y}\delta\omega_{X}\delta\omega_{Y}}{\hbar^{4}}\cos(\phi_{X}-\phi_{Y})\omega_{Z}\zeta_{1}(-\theta)\zeta_{2}(\theta)\frac{\omega_{Z}^{2}(\omega_{X}^{2}+\omega_{Y}^{2})+\omega_{X}^{2}\omega_{Y}^{2}-3\omega_{Z}^4}{\omega_{X}\omega_{Y}(\omega_{X}^{2}-\omega_{Z}^{2})^2(\omega_{Y}^{2}-\omega_{Z}^{2})^{2}}\nonumber\\
\delta_{\pm}&=&-\frac{\omega_{Z}}{\hbar^2}\left(\frac{\zeta_{1}(-\theta)\delta\omega_{X}eF_{X}\exp(\pm\text{i}\phi_{X})}{\omega_{X}\left(\omega_{X}^{2}-\omega_{Z}^{2}\right)}-\frac{\zeta_{2}(-\theta)\delta\omega_{Y}eF_{Y}\exp(\pm\text{i}\phi_{Y})}{\omega_{Y}\left(\omega^{2}_{Y}-\omega_{Z}^{2}\right)}\right)\nonumber\\
&\pm&\text{i}\frac{m^{*}\omega_{Z}}{\hbar^{3}}\left(\exp(\pm\text{i}\phi_{X})\frac{\zeta_{1}(-\theta)\zeta_{2}(\theta)\delta\omega_{X}}{4\omega_{X}^{2}-\omega_{Z}^{2}}+\frac{\exp(\pm\text{i}\phi_{Y})\zeta_{1}(\theta)\zeta_{2}(-\theta)\delta\omega_{Y}}{4\omega_{Y}^{2}-\omega_{Z}^{2}}\right)\nonumber\\
\label{eq:Howdef}
\end{eqnarray}
where $\Delta_{Z}^{12}$ is the higher-order contribution of the
spin-orbit coupling to the energy difference (in Eq.
(\ref{eq:Howdef}), $\Delta_{Z}^{12}$ is only written to second-order
in the spin-orbit coupling), and $\delta_{Z}$ is a Bloch-Siegert
shift\cite{Bloch40} which arises from the noncommutivity of
$\widetilde{H}(t)$ at different times. In the presence of a static
electric field, the coupling between the states $|1\rangle$ and
$|2\rangle$, $\delta_{\pm}$ in Eq. (\ref{eq:Howdef}), is first-order
in the spin-orbit coupling parameters, whereas it is second-order in
the spin-orbit coupling when $F_{X}=F_{Y}=0$ eV/m.

\begin{figure}%[b!]
\includegraphics*[height=14.7cm,width = 12.7cm]{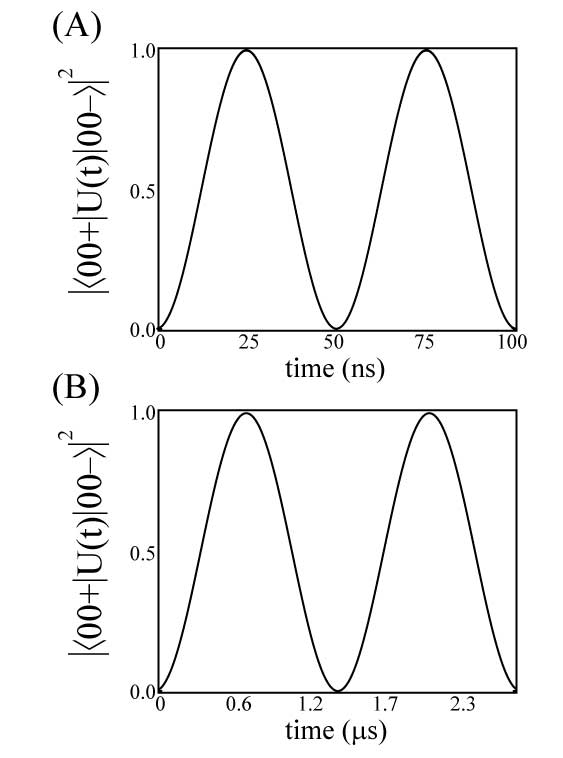}
\caption{Numerical simulation of the transition amplitudes,
$|\langle
0,0,+|T\exp(-\text{i}\int^{t}_{0}\text{d}t'\widetilde{H}(t'))|0,0,-\rangle|^{2}$,
caused by modulating a quantum dot's oscillator frequency,
$\omega_{X}(t)=\omega_{X}+\delta\omega_{X}\sin(\omega_{r}t)$, (A)
with and (B) without a static electric field. In the presence of a
static electric field, the effective Rabi frequency is first-order
in the spin-orbit coupling [Eq. (\ref{eq:Howdef})], which lead to a
large Rabi frequency as shown in Fig. \ref{fig:fig2}(A).  The
following parameters were used in the simulation:
$\hbar\omega_{Y}=\hbar\omega_{X}=1$ meV, $\alpha=4\times 10^{-13}$
eV-m, $\beta=0$ eV-m, $\hbar\omega_{Z}=0.1$ meV, $\theta=0$,
$\delta\omega_{X}=\frac{\omega_{X}}{10}$, $F_{X}=10^{4}$ eV/m, and
$\phi_{X}=0$, which gave a Rabi frequency of 9.86 MHz compared with
the calculated value of $|\delta_{\pm}|/(2\pi)\approx 9.77$ MHz by
Eq. (\ref{eq:Howdef}). In the absence of a static electric field,
the effective Rabi frequency is second-order in the spin-orbit
coupling, which lead to a smaller Rabi frequency as shown in Fig.
\ref{fig:fig2}(B).  The following parameters were used:
 $\hbar\omega_{Y}=1$ meV, $\hbar\omega_{X}=0.25$ meV, $\hbar\omega_{Z}=0.1$ meV,
$\alpha=8\times 10^{-13}$ eV-m, $\delta\omega_{X}=\omega_{X}/20$,
$\beta=0$ eV-m, $\theta=\pi/4$, $F_{X}=0$ eV/m, and $\phi_{X}=0$,
which gave a Rabi frequency of 355 kHz compared with the calculated
value of $|\delta_{\pm}|/(2\pi)=354$ kHz given by Eq.
(\ref{eq:Howdef}).} \label{fig:fig2}
\end{figure}

 Figure
\ref{fig:fig2}(A) presents the exact numerical simulation of the
transition amplitude, $|\langle 0,0,+|U(t,0)|0,0,-\rangle|^2$, for a
parabolic quantum dot under harmonic modulation of $\omega_{X}$ for
$F_{X}\neq 0$. In the simulation, the eigenstates of
$\widetilde{H}_{0}(\omega_{X},\omega_{Y})$ in Eq.
(\ref{eq:Hparmod1}) were numerically found by diagonalizing
$\widetilde{H}_{0}(\omega_{X},\omega_{Y})$ using a basis of four
hundred $|n,m,\pm\rangle$ states. Next, $\widetilde{H}(t)$ in Eq.
(\ref{eq:Hparmod}) was expanded in the $n$ lowest energy eigenstates
of $\widetilde{H}_{0}(\omega_{X},\omega_{Y})$, and
$U(t)=T\exp(-\text{i}/\hbar\int^{t}_{0}\widetilde{H}(t')\text{d}t')$
was then found numerically ($n=30$ was found to give converged
results for the simulations).  The modulation frequency,
$\omega_{r}$, was given by the numerically calculated energy
difference between the two states, $\Delta E_{12}$, plus the Bloch-
Siegert shift given in Eq. (\ref{eq:Howdef}), i.e.,
$\omega_{r}=\Delta E_{12}/\hbar+\delta_{Z}\approx
\omega_{Z}+\Delta_{Z}^{12}+\delta_{Z}$.   This method was used
throughout the paper when numerically calculating the exact
transition amplitudes.

 The following parameters were used in the simulation shown in Fig. \ref{fig:fig2}(A): $\hbar\omega_{Y}=\hbar\omega_{X}=1$
meV, $\hbar\omega_{Z}=0.1$ meV (corresponding to an in-plane
magnetic field of around five Tesla in GaAs), $\alpha=4\times
10^{-13}$ eV-m, $m^{*}=0.067m_{0}$ (where $m_{0}$ is the free
electron mass), $\delta\omega_{X}=\omega_{X}/10$, $F_{X}=10^{4}$
eV/m, and $\theta=0$. With these parameters, the effective Rabi
frequency for ``on-resonance'' modulation [
$(\omega_{r}-\omega_{Z})/(2\pi)=-3.46$ MHz] observed in Fig.
\ref{fig:fig2}(A) was 9.86 MHz, which is consistent with the
theoretical value given by Eq. (\ref{eq:Howdef}),
$|\delta_{\pm}|/(2\pi)=9.76$ MHz. The fact that the Rabi frequency
is directly proportional to the spin-orbit coupling arises because
modulation of $\omega_{X}$ results in a modulation of the effective
center of the quantum dot $\vec{r}_{c}$ by $\delta\vec{r}_{c}$ when
$F_{X}\neq 0$. For the parameters used in Fig. \ref{fig:fig2}(A),
$|\delta \vec{r}_{c}|=2.3$ nm.

For the case when $F_{X}=F_{Y}=0$ eV/m, $\vec{r}_{c}$ remains fixed,
and the electron wave function is only squeezed/expanded during the
modulation [Fig. \ref{fig:fig1}(A)]. Figure \ref{fig:fig2}(B)
presents the exact numerical simulation of the transition amplitude,
$|\langle 0,0,+|U(t,0)|0,0,-\rangle|^2$, when $F_{X}=F_{Y}=0$ eV/m.
 The following parameters were used: $\hbar\omega_{Y}=1$ meV,
$\hbar\omega_{X}=0.25$ meV, $\hbar\omega_{Z}=0.1$ meV,
$\alpha=8\times 10^{-13}$ eV-m, $m^{*}=0.067m_{0}$ (where $m_{0}$ is
the free electron mass), $\delta\omega_{X}=\omega_{X}/20$, and
$\theta=\pi/4$. With these parameters, the effective Rabi frequency
observed in Fig. \ref{fig:fig2}(B) was 355 kHz, which is consistent
with the calculated value given by Eq. (\ref{eq:Howdef}),
$|\delta_{\pm}|/(2\pi)=354$ kHz, for ``on-resonance'' irradiation,
$(\omega_{r}-\omega_{Z})/(2\pi)=-39.285$ MHz.  The observed Rabi
frequency was over an order of magnitude smaller than the case of
nonzero $F_{X}$ [Fig. \ref{fig:fig2}(A)]. As stated earlier, the
effective Rabi frequency is smaller since it is second-order in the
spin-orbit coupling and contains terms like
$(\alpha^{2}-\beta^{2})/2\sin(2\theta)\pm \alpha\beta\cos(2\theta)$.
Note that for $\theta=n\pi/2$, both $\alpha$ and $\beta$ must be
nonzero for $\delta_{\pm}$ to be nonzero. Furthermore, for the case
when $\omega_{X}=\omega_{Y}$ and for uniform modulation (i.e,
$\delta\omega_{Y}=\delta\omega_{X}$), the Rabi frequency in Eq.
(\ref{eq:Howdef}) is exactly proportional to
$\alpha\beta\cos(2\theta)$.  Finally, it should be noted that
$\omega_{r}-\omega_{Z}$ is on the order of $|\delta_{\pm}|$ [Fig.
\ref{fig:fig2}(A)] or much greater than $|\delta_{\pm}|$ [Fig.
\ref{fig:fig2}(B)], so that $\omega_{r}$ must be tuned with
precision given by $|\delta_{\pm}|$ in order to maximize the
amplitude of the Rabi oscillations.

 Although Eq.
(\ref{eq:Hber}) was used to generate $\widehat{H}^{12}_{\text{EFF}}$
in Eqs. (\ref{eq:How})-(\ref{eq:Howdef}), the dynamics under
modulations of $\omega_{X}$ and $\omega_{Y}$ can be calculated using
the following time-dependent Hamiltonian:
\begin{eqnarray}
\widehat{H}(t)&=&\frac{\widehat{P}_{X}^{2}}{2m^{*}}+\frac{\widehat{P}_{Y}^{2}}{2m^{*}}+\widehat{H}_{SO}-\frac{\hbar\omega_{Z}}{2}\widehat{\sigma}_{Z}+\frac{m^{*}}{2}\left(\omega_{X}^{2}(t)\widehat{X}^2+\omega_{Y}^{2}(t)\widehat{Y}^{2}\right)-eF_{X}\widehat{X}-eF_{Y}\widehat{Y}\nonumber\\
\label{eq:hparoo}
\end{eqnarray}
where $\widehat{P}_{X(Y)}=\sqrt{m^*
\hbar\omega_{X(Y)}/2}\text{i}(a^{\dagger}_{X(Y)}-a_{X(Y)})$,
$\widehat{X}(\widehat{Y})=\sqrt{\hbar/(2m^{*}\omega_{X(Y)})}\left(a_{X(Y)}+a^{\dagger}_{X(Y)}\right)$,
and $\omega_{X(Y)}(t)=\omega_{X(Y)}+\delta\omega_{X(Y)}(t)$.
$\widehat{H}(t)$ in Eq. (\ref{eq:hparoo}) can be used to construct
an effective Hamiltonian in the $|0,0,+,1_{F}\rangle$ and
$|0,0,-,0_{F}\rangle$ subspace, which is also given by Eqs.
(\ref{eq:How})-(\ref{eq:Howdef}).
\subsection{Parametric Modulations of a Parabolic Quantum Dot's
center}

\begin{figure}%[b!]
\includegraphics*[height=10.7cm,width = 13.7cm]{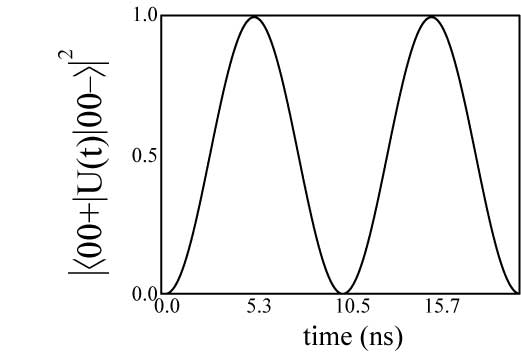}
\caption{Numerical simulation of the transition amplitudes,
$|\langle
0,0,+|T\exp(-\text{i}\int^{t}_{0}\text{d}t'\widetilde{H}(t'))|0,0,-\rangle|^{2}$,
caused by modulating the center of the quantum dot's parabolic
potential.  In this case, the effective Rabi frequency is
first-order in the spin-orbit coupling [Eq. (\ref{eq:d12ef})],
leading to a large Rabi frequency.  The following parameters were
used in the simulation: $\hbar\omega_{Y}=\hbar\omega_{X}=1$ meV,
$\alpha=4\times 10^{-13}$ eV-m, $\beta=0$ eV-m,
$\hbar\omega_{Z}=0.1$ meV, $\theta=0$, $\delta x_{c}=11.5$ nm
[equivalent to using an electric field of $E_{X}=10^{4}$ eV/m], and
$\phi_{X}=0$, which gave a Rabi frequency of 48.7 MHz compared with
calculated value of $|\delta^{\text{EF}}_{\pm}|/(2\pi)\approx 48.8$
MHz given by Eq. (\ref{eq:d12ef}).} \label{fig:fig2a}
\end{figure}

In addition to modulating the confinement frequency, the center of
the parabolic quantum dot can also be modulated, i.e,
$x_{c}\rightarrow x_{c}+\delta x_{c}(t)$ and $y_{c}\rightarrow
y_{c}+\delta y_{c}(t)$. Unlike the case of modulating $\omega_{X}$
and $\omega_{Y}$, displacements of the harmonic potential do not
alter the energy spacings of the quantum dot; however, the
simultaneous eigenstates of the oscillator do change under such
modulation. Using Eq. (\ref{eq:Hber}), the Hamiltonian under
time-dependent displacements of the parabolic potential's center is
given by:
\begin{eqnarray}
\frac{\widetilde{H}(t)}{\hbar}&=&\omega_{X}(a^{\dagger}_{X}a_{X}+1/2)+\omega_{Y}(a^{\dagger}_{Y}a_{Y}+1/2)-\frac{\omega_{Z}}{2}\widehat{\sigma}_{Z}\nonumber\\
&+&\text{i}\sqrt{\frac{m^{*}\omega_{X}}{2\hbar^3}}\left(a^{\dagger}_{X}-a_{X}\right)\left(\zeta_{1}(-\theta)\widehat{\sigma}_{X}+\zeta_{2}(\theta)\widehat{\sigma}_{Z}\right)-\text{i}\sqrt{\frac{m^{*}\omega_{X}}{2\hbar}}\frac{\partial
x_{c}(t)}{\partial t}\left(a^{\dagger}_{X}-a_{X}\right)\nonumber\\
&-&\text{i}\sqrt{\frac{m^{*}\omega_{Y}}{2\hbar^{3}}}\left(a^{\dagger}_{Y}-a_{Y}\right)\left(\zeta_{1}(\theta)\widehat{\sigma}_{Z}+\zeta_{2}(-\theta)\widehat{\sigma}_{X}\right)-\text{i}\sqrt{\frac{m^{*}\omega_{Y}}{2\hbar}}\frac{\partial
y_{c}(t)}{\partial t}\left((a^{\dagger}_{Y})-a_{Y}\right)
\label{eq:centermod}
\end{eqnarray}
For small displacements, $x_{c}\rightarrow x_{c}+\delta
x_{c}\sin(\omega_{r}t+\phi_{X})$ and $y_{c}\rightarrow y_{c}+\delta
y_{c}\sin(\omega_{r}t+\phi_{Y})$, the effective Hamiltonian in the
$|0,0,+,1_{F}\rangle$ and $|0,0,-,0_{F}\rangle$ subspace can be
written as (for $\omega_{r}\approx \omega_{Z}$):
\begin{eqnarray}
\frac{\widehat{H}^{12}_{\text{EFF}}}{\hbar}&=&\frac{1}{2}\left(\omega_{r}-\omega_{Z}+\delta^{c}_{Z}+\Delta^{12}_{Z}\right)\widehat{\sigma}^{12}_{Z}+\delta^{c}_{+}\widehat{\sigma}^{12}_{+}+\delta^{c}_{-}\widehat{\sigma}^{12}_{-}
\end{eqnarray}
where \begin{eqnarray}
\delta^{c}_{Z}&=&\frac{(m^{*})^2\omega_{Z}}{4\hbar^4}\left((\delta
x_{c})^2\frac{\omega^{4}_{X}(\omega_{X}^{2}+3\omega_{Z}^{2})(\zeta_{1}(-\theta))^2}{\left(\omega_{Z}^{2}-\omega_{X}^2\right)^3}+(\delta
y_{c})^{2}\frac{\omega_{Y}^4(\omega_{Y}^{2}+3\omega_{Z}^{2})(\zeta_{2}(-\theta))^{2}}{\left(\omega_{Z}^{2}-\omega_{Y}^{2}\right)^3}\right)\nonumber\\
&+&\frac{(m^{*}\omega_{X}\omega_{Y})^2}{2\hbar^{4}}\cos(\phi^{c}_{X}-\phi^{c}_{Y})\delta
y_{c}\delta
{x}_{c}\omega_{Z}\zeta_{1}(-\theta)\zeta_{2}(\theta)\frac{\omega_{Z}^{2}(\omega_{X}^{2}+\omega_{Y}^{2})+\omega_{X}^{2}\omega_{Y}^{2}-3\omega_{Z}^4}{(\omega_{X}^{2}-\omega_{Z}^{2})^2(\omega_{Y}^{2}-\omega_{Z}^{2})^{2}}\\
\delta^{c}_{\pm}&=&-\frac{m^{*}\omega_{Z}}{2\hbar^{2}}\left(\exp(\pm\text{i}\phi^{c}_{X})\delta
x_{c}\frac{\omega_{X}^{2}\zeta_{1}(-\theta)}{\omega_{Z}^{2}-\omega_{X}^{2}}-\exp(\pm\text{i}\phi^{c}_{Y})\delta
y_{c}\frac{\omega_{Y}^{2}\zeta_{2}(-\theta)}{\omega_{Z}^{2}-\omega_{Y}^{2}}\right)
\label{eq:d12ef}
\end{eqnarray}
The effective coupling between spin states, $\delta_{\pm}^{c}$, is
first-order in the spin-orbit coupling, which leads to large Rabi
frequencies. Figure \ref{fig:fig2a} presents an exact numerical
simulation of the transition amplitude, $|\langle
0,0,+|T\exp(-\text{i}\int^{t}_{0}\text{d}t'\widetilde{H}(t')/\hbar)|0,0,-\rangle|^{2}$,
in the presence of modulating the center of the parabolic well.  The
following parameters were used in the simulation:
$\hbar\omega_{X}=\hbar\omega_{Y}=1$ meV, $\alpha=4\times 10^{-13}$
eV-m, $\beta=0$ eV-m, $\hbar\omega_{Z}=0.1$ meV, $\theta=0$,
$F_{X}=F_{Y}=0$ eV/m, and $\delta x_{c}=11.4 $ nm.  Such parameters
gave an effective Rabi frequency of 48.7 MHz in Fig. \ref{fig:fig2a}
compared with the calculated value of $|\delta^{c}_{+}|/(2\pi)=48.8$
MHz given by Eq. (\ref{eq:d12ef}) for ``on-resonance'' modulation
(i.e., $(\omega_{r}-\omega_{Z})/(2\pi)=-3.435$ MHz). As mentioned in
the introduction, parametric modulation of a parabolic quantum dot's
center is equivalent to applying a time-dependent electric field,
$E(t)=E_{X}(t)\widehat{x}+E_{Y}(t)\widehat{y}$, with the
correspondence that $\delta
x_{c}(t)=eE_{X}(t)/(m^{*}\omega_{X}^{2})$ and $\delta
y_{c}(t)=eE_{Y}(t)/(m^{*}\omega_{Y}^{2})$.  The simulation in Fig.
\ref{fig:fig2a} corresponds to $E_{X}=10^4$ eV/m.  Previous
theoretical work has suggested using such EDSR methods to perform
efficient spin rotations in quantum
wells\cite{Rashba03a,Rashba03,Efros06} and quantum
dots\cite{Golovach06,Flindt06}.
\subsection{Determining the spin-orbit coupling constants, $\alpha$
and $\beta$} In the above examples of parametric oscillations, both
the effective Rabi frequency, $\delta_{\pm}$, and the effective
offset, $\Delta_{Z}$, depend upon the spin-orbit coupling constants
of the quantum dot ($\alpha$ and $\beta$), the oscillator
frequencies ($\omega_{X}$ and $\omega_{Y}$), and the direction of
the magnetic field, $\theta$. This dependence can potentially be
used to determine both the spin-orbit coupling constants $\alpha$
and $\beta$.  For instance, consider the effects of modulating the
center of the parabolic dot, i.e., performing an EDSR experiment.
Taking the ratio of the ``measured'' Rabi frequency [Eq.
(\ref{eq:d12ef})] for an experiment with $\delta
x_{c}=\lambda_{\text{mod}}$ and $\delta y_{c}=0$ to the ``measured''
Rabi frequency for an experiment with $\delta x_c =0$ and $\delta
y_{c}=\lambda_{\text{mod}}$ gives
\begin{eqnarray}
ZZ(\theta)&=&\frac{\delta_{\pm}(\delta
x_{c}=\lambda_{\text{mod}},\delta
y_{c}=0),\theta}{\delta_{\pm}(\delta x_{c}=0,\delta
y_{c}=\lambda_{\text{mod}},\theta)}\nonumber\\&=&-\frac{\zeta_{1}(-\theta)}{\zeta_{2}(-\theta)}
\label{eq:zz}
\end{eqnarray}
For an in-plane magnetic field along the $\widehat{y}$ axis (i.e.,
$\theta=\pi/2$), the relative ratio of the spin-orbit coupling
strengths can be found since in this case $ZZ(\pi/2)=\beta/\alpha$.
Furthermore, $\Delta_{Z}$, which can be determined by tuning
$\omega_{r}$ to the frequency which maximizes the amplitude of the
Rabi oscillations and then approximating
$\Delta_{Z}\approx\omega_{r}-\omega_{Z}$, can then be used to
determine the absolute value of $\alpha$ and $\beta$. Although the
calculations used in this section were for a harmonic potential,
more realistic $\widehat{V}(\widehat{X},\widehat{Y})$ using the full
electrostatic potential generated by the metallic surface
gates\cite{Stopa96} could be used to more accurately characterize
the spin dynamics in terms of $\alpha$ and $\beta$ under parametric
modulation of $\widehat{V}(\widehat{X},\widehat{Y})$.

\section{The effects of the Cubic Dresselhaus spin-orbit
interaction} Recent work has indicated that the cubic Dresselhaus
term can become the dominant spin-orbit interaction in highly
confined (i.e., large $\omega_{X}$ and $\omega_{Y}$) quantum
dots\cite{Jacob07}. For the coordinate system chosen in this work,
the cubic Dresselhaus term is given by:
\begin{eqnarray}
\widehat{H}_{D}^{\text{cub}}&=&\frac{\gamma}{\hbar^3}\left(\widehat{P}_{Y}\widehat{P}_{X}\widehat{P}_{Y}\left(\cos(\theta)\widehat{\sigma}_{Z}-\sin(\theta)\widehat{\sigma}_{X}\right)-\widehat{P}_{X}\widehat{P}_{Y}\widehat{P}_{X}\left(\sin(\theta)\widehat{\sigma}_{Z}+\cos(\theta)\widehat{\sigma}_{X}\right)\right)
\label{eq:cubD}
\end{eqnarray}
The linear Dresselhaus coupling, $\beta$ in Eq. (\ref{eq:Hintro2}),
is related to the cubic Dresselhaus coupling constant, $\gamma$ in
Eq. (\ref{eq:cubD}), by $\beta=\gamma\langle
\widehat{P}^{2}_{Z}\rangle/\hbar^2$ [where for a quantum well of
width $l_{w}$, $\langle\widehat{P}^{2}_{Z}\rangle/\hbar^2\approx
(\pi/l_{w})^{2}$]. For a quantum well with $l_{w}=30$ nm, and for
$\hbar\omega_{X(Y)}\approx 1$ meV, the relative strength of the
cubic and linear Dresselhaus terms is given roughly by
$\langle\widehat{P}^{2}_{X(Y)}\rangle/\langle
\widehat{P}_{Z}^{2}\rangle\approx
m^{*}\omega_{X(Y)}l_{w}^{2}/(2\pi^{2}\hbar)=0.04$. The larger
$\omega_{X}$ and $\omega_{Y}$ are for the quantum dot, the more the
cubic Dresselhaus coupling contributes to the total spin-orbit
interaction.

Under parametric modulation of the center of the quantum dot (i.e.,
an EDSR experiment), the effect of the cubic Dresselhaus coupling
can be found by simply adding  $\widehat{H}^{\text{cub}}_{D}$ to
$\widetilde{H}(t)$ in Eq. (\ref{eq:centermod}). The effective
Hamiltonian in the $|1\rangle=|0,0,+,1_{F}\rangle$ and
$|2\rangle=|0,0,-,0_{F}\rangle$ subspace in this case is given by:
\begin{eqnarray}
\frac{\widehat{H}^{12}_{\text{cub}}}{\hbar}&=&\frac{1}{2}\left(\omega_{r}-\omega_{Z}-\Delta^{12}_{Z,\text{cub}}\right)\widehat{\sigma}^{12}_{Z}+\delta^{c}_{+,\text{cub}}\widehat{\sigma}^{12}_{+}+\delta^{c}_{-,\text{cub}}\widehat{\sigma}^{12}_{-}
\end{eqnarray}
where \begin{eqnarray}
\Delta^{12}_{Z,\text{cub}}&=&\frac{m^{*}\omega_{Z}}{\hbar^3}\left(\frac{\omega_{X}\left(\zeta_{1}^{\text{cub}}(-\theta)\right)^2}{\omega_{Z}^{2}-\omega_{X}^{2}}+\frac{\omega_{Y}\left(\zeta^{\text{cub}}_{2}(-\theta)\right)^{2}}{\omega_{Z}^{2}-\omega_{Y}^{2}}\right)\nonumber\\
&-&\frac{m^{*}\omega_{Z}}{2\hbar}\left(\frac{m^{*}\gamma}{\hbar^{2}}\right)^{2}\omega_{X}\omega_{Y}\left(\frac{\omega_{Y}\sin^{2}(\theta)}{(2\omega_{Y}+\omega_{X})^{2}-\omega_{Z}^{2}}+\frac{\omega_{X}\cos^{2}(\theta)}{(2\omega_{X}+\omega_{Y})^{2}-\omega_{Z}^{2}}\right)\nonumber\\
\delta^{c}_{\pm,\text{cub}}&=&-\frac{m^{*}\omega_{Z}}{2\hbar^{2}}\left(\frac{\exp(\pm\text{i}\phi^{c}_{X})\delta
x_{c}\omega_{X}^{2}\zeta^{\text{cub}}_{1}(-\theta)}{\omega_{Z}^{2}-\omega_{X}^{2}}-\frac{\exp(\pm\text{i}\phi^{c}_{Y})\delta
y_{c}\omega_{Y}^{2}\zeta^{\text{cub}}_{2}(-\theta)}{\omega_{Z}^{2}-\omega_{Y}^{2}}\right)
\label{eq:cub12ef}
\end{eqnarray}
with
$\zeta^{\text{cub}}_{1}(-\theta)=\alpha\cos(\theta)+\left[\beta-\gamma
m^{*}\omega_{Y}/(2\hbar)\right]\sin(\theta)$ and
$\zeta^{\text{cub}}_{2}(-\theta)=-(\alpha\sin(\theta)+\left[\beta-\gamma
m^{*}\omega_{X}/(2\hbar)\right]\cos(\theta))$.  Since $\gamma$ and
$\beta$ have the same sign, the cubic Dresselhaus interaction
lessens the contribution of the linear Dresselhaus interaction to
the effective Rabi frequency under parametric modulation of the
electrostatic potential. Similar results are also obtained in the
case of modulating $\omega_{X}$ and $\omega_{Y}$; in this case, the
effective Rabi frequency is again proportional to the spin-orbit
coupling parameters  only in the presence of a nonzero static
electric field, $F_{X(Y)}\neq 0$ [otherwise, it is second-order in
the spin-orbit coupling parameters].

\begin{figure}%[b!]
\includegraphics*[height=13.2cm,width = 11.4cm]{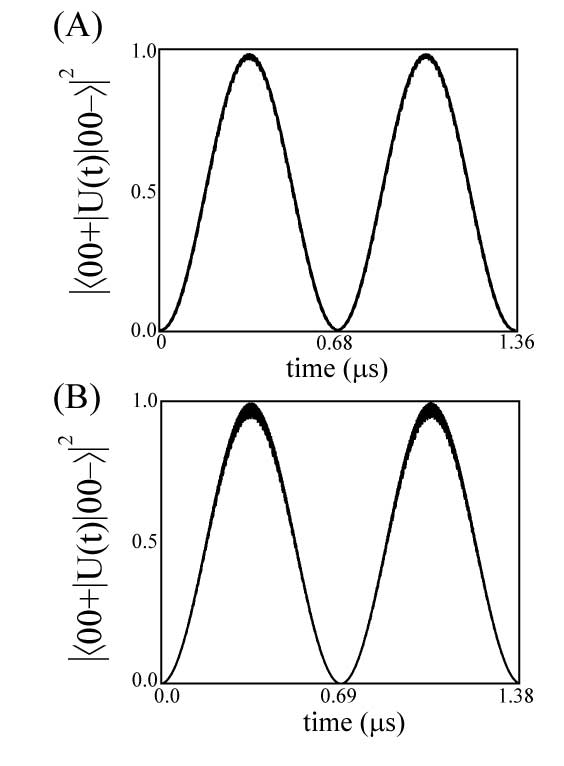}
\caption{Numerical simulation of the transition amplitudes,
$|\langle
0,0,+|T\exp(-\text{i}\int^{t}_{0}\text{d}t'\widetilde{H}(t'))|0,0,-\rangle|^{2}$,
caused by the simultaneous modulation of the quantum dot's
oscillator frequency
[$\omega_{X}(t)=\omega_{X}+\delta\omega_{X}\sin(\omega_{r2}t)$] and
center [$y_{c}(t)=\delta y_{c}\sin(\omega_{r1}t)$], (A) with and (B)
without the Rashba and linear Dresselhaus spin-orbit interaction.
 Efficient spin transitions, which depend upon the cubic
Dresselhaus coupling to first-order, can be performed when
$\omega_{r1}+\omega_{r2}\approx\omega_{Z}$ [Eq. (\ref{eq:rabcub})].
In both simulations, the following parameters were used:
$\hbar\omega_{Y}=0.45$ meV, $\hbar\omega_{X}=3$ meV, ,
$\hbar\omega_{Z}=0.1$ meV, $\theta=0$,
$\delta\omega_{X}=\frac{\omega_{X}}{10}$,
$\omega_{r1}=2\omega_{r2}$, $\delta y_{c}=56$ nm (equivalent to
$E_{Y}=10^4$ eV/m), $\gamma=27\times 10^{-30}$ eV-m$^3$, and
$\phi_{X}=\phi_{Y}=0$. In (A), $\alpha=4\times 10^{-13}$ eV-m and
$l_{w}=30$ nm (giving $\beta=2.96\times 10^{-13}$ eV-m), which gave
a sizable Bloch-Siegert shift of
$\delta^{12}_{Z,\text{cub}}/(2\pi)=3.32$ MHz.  In Fig.
\ref{fig:cub}(A), the observed Rabi frequency was 736.6 kHz, which
is within $2\%$ of the calculated theoretical value given by Eq.
(\ref{eq:rabcub}), $|\delta_{\pm,\text{cub}}^{12}|/(2\pi)=721.5$
kHz. In (B), the linear Dresselhaus and Rashba spin-orbit coupling
constants were artificially set to zero, resulting in
$\delta_{Z,\text{cub}}^{12}/(2\pi)\approx 400$ kHz; the observed
Rabi frequency was 724.5 kHz, which is similar to that observed in
Fig. \ref{fig:cub}(A) and is closer to the calculated value given by
Eq. (\ref{eq:rabcub}).  This demonstrates that the observed Rabi
oscillations are mainly due to the cubic Dresselhaus coupling
[differences between (A) and (B) are mainly due to higher-order
coupling terms involving $\alpha$ and $\beta$].}\label{fig:cub}
\end{figure}

Under parametric modulation of both the quantum dot's center and
oscillator frequency, Rabi oscillations, which depend only upon
$\gamma$, can occur;  measuring the Rabi
frequency\cite{Engel01a,Engel02} would therefore provide an
independent measurement of $\gamma$. Consider the case when the
quantum dot's center is modulated about the $\widehat{y}$ direction
at a frequency of $\omega_{r1}$, $y_{c}(t)=\delta
y_{c}\sin(\omega_{r1}t+\phi_{Y})$, while at the same time the
quantum dot is being periodically squeezed about the $\widehat{x}$
direction at a frequency of $\omega_{r2}$,
$\omega_{X}(t)=\omega_{X}+\delta\omega_{X}\sin(\omega_{r2}t+\phi_{X})$.
In this case, efficient spin rotations can be performed in the
relevant bimodal Floquet subspace,
$|1\rangle=|0,0,+,1_{F1},1_{F2}\rangle$ and
$|2\rangle=|0,0,-,0_{F1},0_{F2}\rangle$ when
$\omega_{r1}+\omega_{r2}\approx \omega_{Z}$ [here $F_{1}$ and
$F_{2}$ denote the Floquet states relative to the oscillator
frequencies $\omega_{r1}$ and $\omega_{r2}$ respectively]. The
effective Hamiltonian in this subspace can be written as:
\begin{eqnarray}
\frac{\widehat{H}^{12}_{\text{cub}}}{\hbar}&=&\frac{1}{2}\left(\omega_{r1}+\omega_{r2}-\omega_{Z}-\Delta^{12}_{Z,\text{cub}}-\delta_{Z,\text{cub}}^{12}\right)\widehat{\sigma}^{12}_{Z}+\delta^{12}_{+,\text{cub}}\widehat{\sigma}^{12}_{+}+\delta^{12}_{-,\text{cub}}\widehat{\sigma}^{12}_{-}
\end{eqnarray}
where \begin{eqnarray} \label{eq:cubbs}
\delta_{Z,\text{cub}}^{12}&\approx&\left(\frac{m^{*}\delta
y_{c}}{\hbar^2}\right)^2\frac{\omega_{r1}^{2}\omega_{Y}^{4}\omega_{Z}\left(\zeta^{\text{cub}}_{2}(-\theta)\right)^{2}}{\left(\omega_{r1}^2-\omega_{Y}^{2}\right)^2\left(\omega_{Z}^{2}-\omega_{r1}^{2}\right)}\\
\delta_{\pm,\text{cub}}^{12}&=&\pm\frac{\text{i}\gamma\cos(\theta)}{2\hbar^3}\exp(\pm\text{i}(\phi_{X}+\phi_{Y}))\frac{\delta\omega_{X}\delta
y_{c}\omega_{r1}\left(m^{*}\omega_{X}\omega_{Y}\right)^{2}}{(\omega_{r1}^{2}-\omega_{Y}^{2})(\omega_{r2}^{2}-4\omega_{X}^2)}
\label{eq:rabcub}
\end{eqnarray}
The effect of the linear Dresselhaus and Rashba spin-orbit coupling
arises mainly in the Bloch-Siegert shift,
$\delta_{Z,\text{cub}}^{12}$, and in $\Delta_{Z,\text{cub}}^{12}$;
higher-order terms contributions to $\delta_{\pm,\text{cub}}^{12}$
arising from both the linear Dresselhaus and Rashba spin-orbit
coupling can be made negligible by using smaller $\delta\omega_{X}$
and $\delta y_{c}$.

Figure \ref{fig:cub} shows the numerical simulation of $|\langle
0,0,+|T\exp(-\frac{i}{\hbar}\int^{t}_{0}\widetilde{H}(t')\text{d}t')|0,0,-\rangle|^{2}$
under both periodic modulation of the quantum dot's center about the
$\widehat{y}$ direction and modulation of the oscillator frequency
about the $\widehat{x}$ direction in (A) the presence of and in (B)
the absence of the Rashba and linear Dresselhaus spin-orbit
interactions. In Fig. \ref{fig:cub}(A), the following parameters
were used: $\hbar\omega_{X}=3$ meV, $\hbar\omega_{Y}=0.45$ meV,
$\hbar\omega_{Z}=0.1$ meV, $\delta\omega_{X}=\omega_{X}/10$,
$\omega_{r1}=2\omega_{r2}$, $\gamma=27\times 10^{-30}$ eV-m$^3$,
$\alpha=4\times 10^{-13}$ eV-m, $l_{w}=30$ nm (giving $\beta\approx
2.96\times 10^{-13}$ eV-m), and $\delta y_{c}=56$ nm (giving
$E_{Y}\approx 10^{4}$ eV/m).  With the above parameters,
$\delta_{Z,\text{cub}}^{12}/(2\pi)\approx 3.33$ MHz.  The observed
Rabi frequency in Fig. \ref{fig:cub}(A) was 736.6 kHz, which is
within $2\%$ of the calculated value given by Eq. (\ref{eq:rabcub}),
$|\delta_{\pm,\text{cub}}^{12}|/(2\pi)=721.5$ kHz.  Since
$\delta^{12}_{Z,\text{cub}}>|\delta^{12}_{\pm,\text{cub}}|$,
$\omega_{r1}$ and $\omega_{r2}$ must be tuned to include
$\delta^{12}_{Z,\text{cub}}$ in order to maximize the amplitude of
the Rabi oscillations [in Fig. \ref{fig:cub}(A),
$(\omega_{r1}+\omega_{r2}-\omega_{Z})/(2\pi)=-4.504$
MHz+$\delta_{Z,\text{cub}}^{12}/(2\pi)$].  Note also that in the
numerical simulation of Fig. \ref{fig:cub}(A), the transition
amplitude does not go exactly to 1 (maximum value of 0.9861), which
is most likely due to higher-order corrections to the Bloch-Siegert
shift, $\delta^{12}_{Z,\text{cub}}$.

 Better agreement between the
Rabi frequency calculated using Eq. (\ref{eq:rabcub}) to numerical
simulation is obtained in Fig. \ref{fig:cub}(B) where both $\alpha$
and $\beta$ were artificially set to zero, removing the dominant
higher-order corrections to both $\delta_{Z,\text{cub}}^{12}$ and
$\delta_{\pm,\text{cub}}^{12}$. In Fig. \ref{fig:cub}(B), the
observed Rabi frequency was 724.5 kHz, which is closer to the
calculated value given by Eq. (\ref{eq:rabcub}),
$|\delta_{\pm,\text{cub}}^{12}|/(2\pi)=721.5$ kHz.  It should be
noted that both simulations shown in Fig. \ref{fig:cub} give roughly
the same Rabi frequency, which, to a good approximation, is given by
Eq. (\ref{eq:rabcub}).  Thus measuring the Rabi frequency under such
parametric modulations should enable $\gamma$ to be directly
determined. It should be noted that for a quantum dot with a
non-harmonic potential and in the presence of an out of plane
magnetic field, it has previously\cite{Golovach06} been found that
the EDSR technique also generates spin rotations which, to first
order, depend upon $\gamma$ and the cyclotron frequency.

\section{Parametric modulations in square quantum dots}
Besides harmonic potentials, another model electrostatic potential
for electrons in lateral quantum dots is that of a square-box (hard
wall) potential defined by $V(x,y)=0$ for $0\leq x\leq L_{X}$ and
$0\leq y\leq L_{Y}$ and $V(x,y)=\infty$ everywhere else. Although
the hard wall potential is not very realistic in two-dimensional
systems, previous studies have utilized such models in analyzing
transport in quantum dots\cite{Rasanen03,Bryant87}, and such
potentials are the quintessential model for studies of chaos in
two-dimensional systems, such as in the stadium
billiard\cite{Tomsovic91,Marcus93}. For the square-box potential,
the eigenstates for $\widehat{H}_{0}$ are simply the two-dimensional
particle in a box states,
$\langle\vec{r}|n,m,\pm\rangle=\sqrt{\frac{2}{L_{X}}}\sqrt{\frac{2}{L_{Y}}}\sin\left(\frac{\pi
nx}{L_{X}}\right)\sin\left(\frac{\pi m y}{L_{Y}}\right)|\pm\rangle$.
In order to induce transitions between the various
$|n,m,\pm\rangle$, small, periodic modulations in both the length
and the width of the box, $L_{X}$ and $L_{Y}$, can be made.  Using
Eq. (\ref{eq:Hber}), the effective time-dependent Hamiltonian during
modulation of the box's length and width, $L_{X}\rightarrow
L_{X}(t)$ and $L_{Y}\rightarrow L_{Y}(t)$, can be written as:
\begin{eqnarray}
\frac{\widetilde{H}(t)}{\hbar}&=&\frac{\hbar\pi^{2}}{2m^{*}L^{2}_{X}(t)}\widehat{H}_{X}+\frac{\hbar\pi^{2}}{2m^{*}L^{2}_{Y}(t)}\widehat{H}_{Y}-\frac{\omega_{Z}}{2}\widehat{\sigma}_{Z}\nonumber\\
&+&\frac{L_{X}}{\hbar^{2}
L_{X}(t)}\widehat{P}_{X}\left(\zeta_{1}(-\theta){\sigma}_{X}+\zeta_{2}(\theta)\widehat{\sigma}_{Z}\right)-\frac{L_{Y}}{\hbar
L_{Y}(t)}\widehat{P}_{Y}\left(\zeta_{1}(\theta)\sigma_{Z}+\zeta_{2}(-\theta)\widehat{\sigma}_{X}\right)\nonumber\\
&+&\frac{\partial L_{X}(t)}{\partial t}\frac{1}{2\hbar
L_{X}(t)}\left(\widehat{P}_{X}\widehat{X}+\widehat{X}\widehat{P}_{X}\right)+\frac{\partial
L_{Y}(t)}{\partial t}\frac{1}{2\hbar
L_{Y}(t)}\left(\widehat{P}_{Y}\widehat{Y}+\widehat{Y}\widehat{P}_{Y}\right)
\label{eq:Hboxmod}
\end{eqnarray}
where \begin{eqnarray}
\widehat{H}_{X}&=&\sum_{n,m,\pm}n^{2}|n,m,\pm\rangle\langle
n,m,\pm|\nonumber\\
\widehat{H}_{Y}&=&\sum_{n,m,\pm}m^{2}|n,m,\pm\rangle\langle n,m,\pm|
\end{eqnarray}
The form of $\widetilde{H}(t)$ arises naturally from Eq.
(\ref{eq:Hber});  previous work on parametric deformations of hard
wall potentials have also arrived at a similar form for the
effective Hamiltonian\cite{Seba90,Razavy91}.  Note, however, that it
is necessary to separate the transformation $\widehat{W}(t)$ from
the dynamics and only include its effect at the end of the
calculation\cite{Cohen01}.
\begin{figure}%[b!]
\includegraphics*[height=14.7cm,width = 12.7cm]{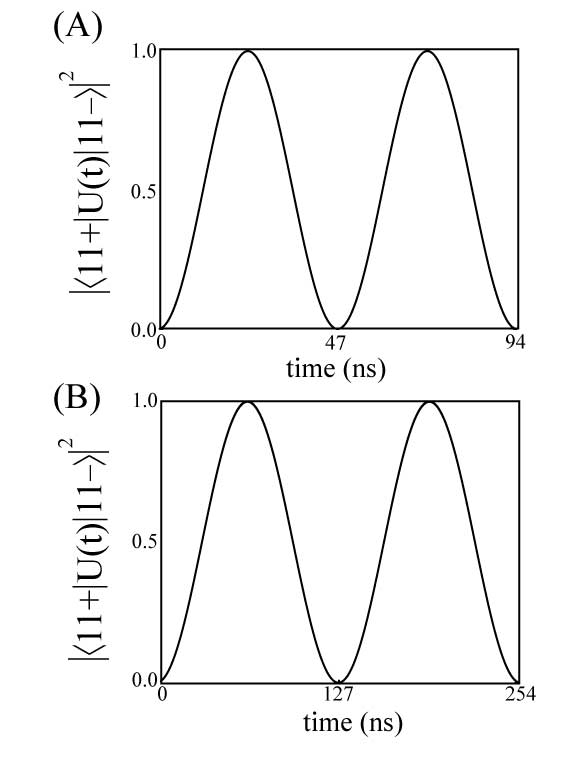}
\caption{Numerical simulation of the transition amplitudes,
$|\langle
1,1,+|T\exp(-\text{i}\int^{t}_{0}\text{d}t'\widetilde{H}(t'))|1,1,-\rangle|^{2}$,
in a square quantum dot caused by either (A) modulating the size of
the quantum dot or (B) by applying a time-dependent electric field.
In both cases, the Rabi frequency is first-order in the spin-orbit
coupling strength. The parameters used in both simulations were
$L_{X}=L_{Y}=70$ nm [giving $\hbar^2\pi^2/(2m^{*}L_{X(Y)}^{2})=1.1$
meV], $\alpha=4\times 10^{-13}$ eV-m, $\beta=0$ eV-m,
$\hbar\omega_{Z}=0.1$ meV, $\theta=0$,
$(\omega_{r}-\omega_{Z})/(2\pi)=-958.4$ kHz, and $\phi_{X}=0$. In
(A), $L_{X}(t)=L_{X}+\delta L_{X}\sin(\omega_{r}t)$ with $\delta
L_{X}=5$ nm, which gave an effective Rabi frequency of 10.6 MHz,
which is in excellent agreement with the calculated value given by
Eq. (\ref{eq:rabbox}), $|\delta_{\pm}|/(2\pi)=10.6$ MHz. In (B), the
applied electric field was
$\vec{E}(t)=E_{X}\sin(\omega_{r}t)\widehat{x}$ with $E_{X}=10^{4}$
eV/m, which gave an effective Rabi frequency of 3.94 MHz, which is
in excellent agreement with the calculated value given by Eq.
(\ref{eq:boxef}), $|\delta_{\pm}^{\text{EF}}|/(2\pi)=3.92$ MHz.}
\label{fig:fig4}
\end{figure}

By performing small modulations of the length and width,
$L_{X}\rightarrow L_{X}+\delta L_{X}(t)$ and $L_{Y}\rightarrow L_{Y}
+\delta L_{Y}(t)$, spin transitions can be performed due to the
spin-orbit interaction.  The effective Hamiltonian in the
$|1\rangle\equiv |1,1,+,1_{F}\rangle$ and $|2\rangle\equiv
|1,1,-,0_{F}\rangle$ Floquet subspace is given to first-order in
$\delta L_{X(Y)}(t)$ and for $\omega_{r}\approx \omega_{Z}$ as:

\begin{eqnarray}
\frac{\widehat{H}^{12}_{\text{EFF}}}{\hbar}&=\frac{1}{2}\left(\omega_{r}-\omega_{Z}+\Delta_{Z}^{12}\right)\widehat{\sigma}^{12}_{Z}+\delta_{+}\widehat{\sigma}^{12}_{+}+\delta_{-}\widehat{\sigma}^{12}_{-}
\label{eq:h12box}
\end{eqnarray}
where
\begin{eqnarray}
\Delta_{Z}^{12}&=&\frac{32m^{*}}{\pi^{2}\hbar^{3}}\sum_{n=2}^{\infty}\frac{n^2\left(1+(-1)^{n}\right)}{\left(n^{2}-1\right)^{2}}\left(\frac{\left(\zeta_{1}(-\theta)\right)^{2}\eta_{X}}{\left(n^{2}-1\right)^{2}-(\eta_{X})^{2}}+\frac{\left(\zeta_{2}(-\theta)\right)^{2}\eta_{Y}}{\left(n^{2}-1\right)^{2}-(\eta_{Y})^{2}}\right)\\
\delta_{\pm}&=&\frac{4}{\hbar}\sum_{n=2}^{\infty}\frac{n^2\left(1+(-1)^{n}\right)}{1-n^2}\left(\frac{\zeta_{1}(-\theta)\eta_{X}\delta
L_{X}\exp(\pm\text{i}\phi_{X})}{L_{X}^{2}\left(\left(n^{2}-1\right)^2-\eta_{X}^2\right)}-\frac{\zeta_{2}(-\theta)\eta_{Y}\delta
L_{Y}\exp(\pm\text{i}\phi_{Y})}{L_{Y}^{2}\left(\left(n^{2}-1\right)^2-\eta_{Y}^2\right)}\right)\nonumber\\\label{eq:rabbox}
\end{eqnarray}
where $\eta_{X(Y)}=2m^{*}\omega_{Z}L_{X(Y)}^{2}/(\hbar\pi^{2})$. By
choosing $\omega_{r}=\omega_{Z}-\Delta^{12}_{Z}$, efficient spin
transitions can be generated in square quantum dots [Fig.
\ref{fig:fig4}(A)] since the Rabi frequency,
$|\delta_{\pm}|/(2\pi)$, is directly proportional to the spin-orbit
coupling strength, in contrast to a parabolic quantum dot undergoing
modulations of confining frequency and in the absence of a static
electric field [Eq. (\ref{eq:Howdef})]. Figure \ref{fig:fig4}(A)
presents an exact numerical simulation of the transition amplitude,
$|\langle
1,1,+|T\exp(-\text{i}\int^{t}_{0}\text{d}t'\widetilde{H}(t')/\hbar)|1,1,-\rangle|^{2}$,
in the presence of modulating the effective size of a square quantum
dot. The following parameters were used in the simulation:
$L_{X}=L_{Y}=70$ nm [giving $\hbar^2 \pi^2/(2m^{*}L_{X(Y)}^2)\approx
1.1$ meV and $\eta_{X(Y)}=0.087$], $\delta L_{X}=5$ nm,
$\alpha=4\times 10^{-13}$ eV-m, $\beta=0$ eV-m,
$\hbar\omega_{Z}=0.1$ meV, $\theta=0$, and
$(\omega_{r}-\omega_{Z})/(2\pi)=-958.4$ kHz. Such parameters gave an
effective Rabi frequency of 10.6 MHz, which is in excellent
agreement with the calculated value given by Eq. (\ref{eq:rabbox}),
$|\delta_{+}|/(2\pi)=10.6$ MHz.

Alternatively, an electric field,
$\vec{E}(t)=E_{X}\sin(\omega_{r}t+\phi_{X})\widehat{x}+E_{Y}\sin(\omega_{r}t+\phi_{Y})\widehat{y}$,
applied to the quantum dot can also induce single spin rotations by
using EDSR effects.  Incorporating the interaction with the electric
field, $\widehat{V}(t)=-e\vec{E}(t)\cdot\vec{r}$, an effective
Hamiltonian in the $|1\rangle$ and $|2\rangle$ subspace [when
$\omega_{r}\approx \omega_{Z}$] can be written as:
\begin{eqnarray}
\frac{\widehat{H}^{12}_{\text{EFF}}}{\hbar}&=\frac{1}{2}\left(\omega_{r}-\omega_{Z}+\Delta_{Z}^{12}\right)\widehat{\sigma}^{12}_{Z}+\delta^{\text{EF}}_{+}\widehat{\sigma}^{12}_{+}+\delta^{\text{EF}}_{-}\widehat{\sigma}^{12}_{-}
\label{eq:h12efbox}
\end{eqnarray}
where
\begin{eqnarray}
\label{eq:boxef}
\delta^{\text{EF}}_{\pm}&=&\frac{32m^{*}}{\pi^{4}\hbar^{3}}\sum_{n=2}^{\infty}\frac{n^2\left(1+(-1)^{n}\right)}{\left(n^2-1\right)^{3}}\left(\frac{L_{X}^{2}E_{X}\zeta_{1}(-\theta)\eta_{X}\exp(\pm\text{i}\phi_{X})}{\left(n^{2}-1\right)^2-\eta_{X}^2}-\frac{L_{Y}^{2}E_{Y}\zeta_{2}(-\theta)\eta_{Y}\exp(\pm\text{i}\phi_{Y})}{\left(n^{2}-1\right)^2-\eta_{Y}^2}\right)\nonumber\\
\end{eqnarray}
Like the case of modulating the size of the quantum dot, the
effective Rabi frequency, $|\delta^{\text{EF}}_{\pm}|/(2\pi)$, is
again proportional to the spin-orbit coupling strength.  Figure
\ref{fig:fig4}(B) presents an exact numerical simulation of the
transition amplitude, $|\langle
1,1,+|T\exp(-\text{i}\int^{t}_{0}\text{d}t'\widetilde{H}(t')/\hbar)|1,1,-\rangle|^{2}$,
in the presence of an electric field,
$\vec{E}(t)=E_{X}\sin(\omega_{r}t+\phi_{X})\widehat{x}$. The
following parameters were used in the simulation: $L_{X}=L_{Y}=70$
nm [giving $\hbar^2 \pi^2/(2m^{*}L_{X(Y)}^2)\approx 1.1$ meV and
$\eta_{X(Y)}=0.087$], $E_{X}=10^4$ eV/m, $\alpha=4\times 10^{-13}$
eV-m, $\beta=0$ eV-m, $\hbar\omega_{Z}=0.1$ meV, $\theta=0$, and
$(\omega_{r}-\omega_{Z})/(2\pi)=-958.4$ kHz. Such parameters gave an
effective Rabi frequency of 3.94 MHz, which is in excellent
agreement with the calculated value given by Eq. (\ref{eq:boxef}),
$|\delta^{\text{EF}}_{+}|/(2\pi)=3.92$ MHz.
%\begin{figure}%[b!]
%\includegraphics{parabmod}
%\caption{By changing the confinement potential, the wave function,
%$\Psi(\omega)$, also changes.  If the frequency of the modulation is
%resonant with the energy difference between two levels of the
%system, transitions can occur.} \label{fig:fig1}
%\end{figure}

\section{Parametric Orbital and Spin Excitations for Increasing spin polarization}
\begin{figure}%[b!]
\includegraphics*[height=14.7cm,width = 12.7cm]{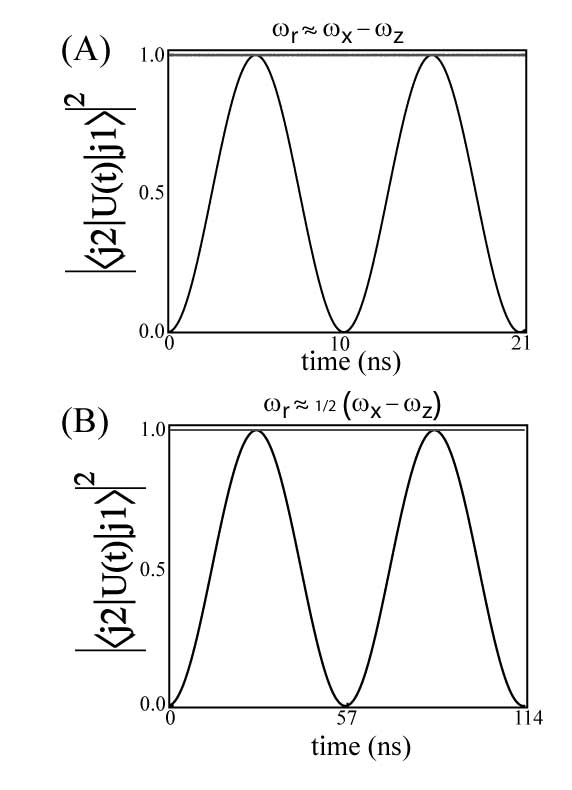}
\caption{(Color online)  Numerical simulation of the transition
amplitudes, $|\langle
0,0,-|T\exp(-\text{i}\int^{t}_{0}\text{d}t'\widetilde{H}(t'))|1,0,+\rangle|^{2}$
(black curve) and $|\langle
0,0,+|T\exp(-\text{i}\int^{t}_{0}\text{d}t'\widetilde{H}(t'))|0,0,+\rangle|^{2}$
(red curve) caused by modulating the oscillator strength defining
the parabolic quantum dot, when (A) $\omega_{r}\approx
\omega_{X}-\omega_{Z}$ and (B) $\omega_{r}\approx
(\omega_{X}-\omega_{Z})/2$. The following parameters
 where used in the simulations:  $\hbar\omega_{Y}=1$ meV, $\hbar\omega_{X}=0.25$ meV,
 $\alpha=4\times 10^{-13}$ eV-m, $\beta=0$ eV-m, $\hbar\omega_{Z}=0.1$
meV, $\theta=0$, $\delta\omega_{X}=\frac{\omega_{X}}{10}$, and
$\phi_{X}=0$.  In (A), modulation of $\omega_{X}$ at a frequency of
$\omega_{r}/(2\pi)\approx (\omega_{X}-\omega_{Z})/(2\pi)=36.3$ GHz
can generate effective transitions between the states
$|0,0,-\rangle$ and $|1,0,+\rangle$, which gave a Rabi frequency of
48.1 MHz, which is in excellent agreement with the calculated value
given by Eq. (\ref{eq:Ho2def}), $|\delta^{23}_{+}|/(2\pi)=48.3$ MHz.
In (B), modulation of $\omega_{X}$ at a frequency of
$\omega_{r}/(2\pi)\approx (\omega_{X}-\omega_{Z})/(4\pi)=18.15$ GHz
resulted in a Rabi frequency of 8.77 MHz, which is within $6\%$ of
the calculated value given in Eq. (\ref{eq:Hoefdef2}),
$|\delta_{2\omega,\pm}^{23}|/(2\pi)\approx 8.26$ MHz. Although the
resulting Rabi frequency is smaller in Fig. \ref{fig:fig3}(B) than
in Fig. \ref{fig:fig3}(A), the required $\omega_{r}$ is a factor of
two smaller, which makes such modulations more experimentally
feasible.}\label{fig:fig3}
\end{figure}

 In addition to single spin
manipulations in parabolic quantum dots, combined transitions
between both the spin and the orbital degrees of freedom can be
generated under parametric modulation of
$\widehat{V}(\widehat{X},\widehat{Y})$. In a parabolic quantum dot,
application of an electric field [$\widetilde{H}(t)$ in Eq.
(\ref{eq:centermod})] and/or modulation of $\omega_{X}$ and
$\omega_{Y}$ [$\widetilde{H}(t)$ in Eq. (\ref{eq:Hparmod1})] can
directly induce both orbital and spin excitations. Such transitions
are an essential component for proposals to increase spin
polarization in a quantum dot's ground electronic
state\cite{Friesen04}.  Consider first the case of parametrically
modulating a parabolic quantum dot's oscillator frequency,
$\omega_{X}$, in the absence of any static electric fields
[$\widetilde{H}(t)$ in in Eq. (\ref{eq:Hparmod1}) with
$F_{X}=F_{Y}=0$ eV/m]. In the following,we will neglect the cubic
Dresselhaus interaction [$\widehat{H}_{D}^{\text{cub}}$ in Eq.
(\ref{eq:cubD})] and will take $\omega_{Y}>\omega_{X}$ in order that
the parametric modulation can be chosen to connect states involving
changes in only one of the orbital degrees of freedom, such as
$|0,0\rangle$ going to $|1,0\rangle$. If we wish to generate
transitions between the states $|1,0,+\rangle$ and $|0,0,-\rangle$,
$\omega_{X}$ must be modulated at frequency of $\omega_{r}\approx
\omega_{X}-\omega_{Z}$. Labeling the relevant states as $|2\rangle =
|0,0,-,1_{F}\rangle$ and $|3\rangle=|1,0,+,0_{F}\rangle$, the
effective Hamiltonian in the $\{|2\rangle, |3\rangle\}$ Floquet
subspace under parametric modulation of $\omega_{X}$ (at
$\omega_{r}\approx \omega_{X}-\omega_{Z}$) is given by:
\begin{eqnarray}
\frac{\widehat{H}^{23}_{\text{EFF}}}{\hbar}&=&\frac{1}{2}\left(\omega_{Z}-\omega_{X}+\omega_{r}+\Delta^{23}_{Z}+\delta^{23}_{Z}\right)\widehat{\sigma}^{23}_{Z}+\delta^{23}_{+}\widehat{\sigma}^{23}_{+}+\delta^{23}_{-}\widehat{\sigma}^{23}_{-}
\label{eq:Hoefa}
\end{eqnarray}
\begin{eqnarray}
\Delta^{23}_{Z}&=&\frac{m^{*}\omega_{Z}}{\hbar^{3}}\left(\frac{2\omega_{X}\left(\zeta_{1}(-\theta)\right)^{2}}{\omega_{Z}^{2}-\omega^{2}_{X}}+\frac{\omega_{Y}\left(\zeta_{2}(-\theta)\right)^{2}}{\omega_{Z}^{2}-\omega_{Y}^{2}}\right)\nonumber\\
\delta^{23}_{Z}&\approx&\frac{\delta\omega_{X}^2}{4\omega_{X}}\left(\frac{(\omega_{X}-\omega_{Z})^2}{4\omega_{X}^{2}-(\omega_{X}-\omega_{Z})^2}\right)\nonumber\\
\delta^{23}_{\pm}&=&\sqrt{\frac{m\omega_{X}}{2\hbar^{3}}}\frac{\omega_{Z}\delta\omega_{X}\zeta_{1}(-\theta)}{\omega_{X}^{2}-\omega_{Z}^{2}}\exp(\pm\text{i}\phi_{X})
\label{eq:Ho2def}
\end{eqnarray}
By choosing
$\omega_{r}=\omega_{X}-\omega_{Z}-\Delta^{23}_{Z}-\delta^{23}_{Z}$,
efficient spin and orbital transitions between the states
$|2\rangle$ and $|3\rangle$ can be generated [Fig.
\ref{fig:fig3}(A)] since the Rabi frequency, $|\delta_{\pm}^{23}|$,
is directly proportional to the spin-orbit coupling. Note that there
can be a significant Bloch-Siegert shift, $\delta_{Z}^{23}$, since
the transition involves different electronic orbitals.

 Figure \ref{fig:fig3}(A) presents an exact numerical simulation of
the transition amplitudes, $|\langle
1,0,+|T\exp(-\text{i}\int^{t}_{0}\text{d}t'\widetilde{H}(t')/\hbar)|0,0,-\rangle|^{2}$
(black curve) and $|\langle
0,0,+|T\exp(-\text{i}\int^{t}_{0}\text{d}t'\widetilde{H}(t')/\hbar)|0,0,+\rangle|^{2}$
(red curve), in the presence of the modulation
$\omega_{X}(t)=\omega_{X}+\delta\omega_{X}\sin(\omega_{r}t)$. The
following parameters were used in the simulation:
$\hbar\omega_{Y}=1$ meV, $\hbar\omega_{X}=0.25$ meV, $\alpha=4\times
10^{-13}$ eV-m, $\beta=0$ eV-m, $\hbar\omega_{Z}=0.1$ meV,
$\theta=0$, $F_{X}=F_{Y}=0$ eV/m, and
$\delta\omega_{X}=\omega_{X}/10$. Such parameters gave an effective
Rabi frequency 48.1 MHz, which is in excellent agreement with the
calculated value given by Eq. (\ref{eq:Ho2def}),
$|\delta^{23}_{+}|/(2\pi)=48.3$ MHz, under ``on-resonant''
modulation [$(\omega_{r}-(\omega_{X}-\omega_{Z}))/(2\pi)=32.32$ MHz
$-\delta_{Z}^{23}/(2\pi)=17.35$ MHz]. Note that the state
$|0,0,+\rangle$ essentially does not evolve into any of the other
states [the red line in Fig. \ref{fig:fig3}(A)] thus confirming that
the electron must be spin down in order to be excited from the state
$|0,0\rangle$. As mentioned earlier, such spin selective transitions
can be used to increase the spin polarization of a quantum
dot\cite{Friesen04}, which will be
 discussed later in this section.

 Since the relevant transitions involve both orbital and spin
 excitation, the required modulation frequency is quite large,
$\omega_{r}\approx \omega_{X}-\omega_{Z}\approx 36.3$ GHz, which can
be experimentally challenging.  However, higher-order processes
which utilize smaller values of $\omega_{r}$ can be found in order
to induce efficient spin and orbital transitions. The simplest
higher-order process to induce transitions between the states
$|1,0,+\rangle$ and $|0,0,-\rangle$ occurs when $\omega_{r}\approx
(\omega_{X}-\omega_{Z})/2$. In this case, the Floquet states
$|1,0,+,0_{F}\rangle$ and $|0,0,-,2_{F}\rangle$ are degenerate. The
effective Hamiltonian in this subspace is given by [for
$\omega_{r}\approx (\omega_{X}-\omega_{Z})/2$]:
\begin{eqnarray}
\frac{\widehat{H}_{2\omega}^{23}}{\hbar}&=&\frac{1}{2}\left(\omega_{Z}-\omega_{X}+2\omega_{r}+\Delta^{23}_{Z}+\delta^{23}_{2\omega,Z}\right)\widehat{\sigma}^{23}_{Z}+\delta^{23}_{2\omega,+}\widehat{\sigma}^{23}_{+}+\delta^{23}_{2\omega,-}\widehat{\sigma}^{23}_{-}
\label{eq:Hoefb}
\end{eqnarray}
where
\begin{eqnarray}
\delta_{2\omega,Z}^{23}&\approx&\frac{\delta\omega_{X}^{2}}{4\omega_{X}}\left(\frac{(\omega_{X}-\omega_{Z})^{2}}{16\omega_{X}^{2}-(\omega_{X}-\omega_{Z})^{2}}\right)\nonumber\\
\delta_{2\omega,\pm}^{23}&=&\mp\text{i}\sqrt{\frac{m^{*}}{32\omega_{X}\hbar^3}}\frac{\left(13\omega_{X}^2+2\omega_{X}\omega_{Z}+\omega_{Z}^2\right)\delta\omega_{X}^{2}\zeta_{1}(-\theta)}{(\omega_{X}-\omega_{Z})^2(\omega_{X}+\omega_{Z})(3\omega_{X}+\omega_{Z})}\exp\left(\pm\text{i}2\phi_{X}\right)
\label{eq:Hoefdef2}
%\left(\frac{7\omega_{X}^{4}+98\omega_{X}\omega_{Z}+8\omega_{X}^{2}\omega_{Z}^{2}+14\omega_{X}\omega_{Z}^{3}+\omega_{Z}^{4}}{32\omega_{X}^{2}(\omega_{X}-\omega_{Z})^{2}(3\omega_{X}^2+4\omega_{X}\omega_{Z}+\omega_{Z}^{2})}\right)\exp\left(\pm
%2\text{i}\phi_{X}\right)
\end{eqnarray}
In this case, the effective Rabi frequency is again directly
proportional to the spin-orbit coupling strength and is also
proportional to the square of the modulation strength,
$(\delta\omega_{X})^{2}$.

Figure \ref{fig:fig3}(B) presents an exact numerical simulation of
the transition amplitudes, $|\langle
1,0,+|T\exp(-\text{i}\int^{t}_{0}\text{d}t'\widetilde{H}(t')/\hbar)|0,0,-\rangle|^{2}$
(black curve) and $|\langle
0,0,+|T\exp(-\text{i}\int^{t}_{0}\text{d}t'\widetilde{H}(t')/\hbar)|0,0,+\rangle|^{2}$
(red curve), in the presence of modulating the oscillator frequency
of the parabolic well at
$\omega_{r}\approx(\omega_{X}-\omega_{Z})/2$. The following
parameters were used in the simulation: $\hbar\omega_{Y}=1$ meV,
$\hbar\omega_{X}=0.25$ meV, $\alpha=4\times 10^{-13}$ eV-m,
$\beta=0$ eV-m, $\hbar\omega_{Z}=0.1$ meV, $\theta=0$,
$\delta\omega_{X}=\omega_{X}/10$, and
$(\omega_{r}-1/2(\omega_{X}-\omega_{Z}))/(2\pi)=-\delta_{2\omega,Z}^{23}/(4\pi)+16.16$
MHz = 14.42 MHz. The above parameters gave an effective Rabi
frequency of 8.77 MHz, which is within $6\%$ of the calculated value
given by Eq. (\ref{eq:Hoefdef2}), $|\delta^{23}_{+}|/(2\pi)\approx
8.26$ MHz. Note also that the state $|0,0,+\rangle$ doesn't evolve
during the parametric modulation of $\omega_{X}$ [red curve in Fig.
\ref{fig:fig3}(B)].

In addition to modulating the confinement frequency, the spin-orbit
coupling constants, $\alpha$ and $\beta$, can also be modulated in
order to produce combined spin and orbital excitations which are
first-order in $\alpha$ and $\beta$. By modulating an electric field
generated by a surface gate above the quantum dot, the Rashba
spin-orbit coupling parameter within the quantum dot can be
controlled\cite{Nitta97}; additionally, for quantum dots formed
using semiconductor heterostructures, surface gates could also be
used to change $\langle \widehat{P}_{Z}^{2}\rangle$ thereby changing
the linear Dresselhaus coupling constant, $\beta=\gamma\langle
\widehat{P}_{Z}^{2}\rangle$. Such modulations of the spin-orbit
interaction have been previously suggested as a way to perform both
spin and orbital excitations in the absence of a magnetic
field\cite{Debald05}. Finally, it should be noted that application
of an electric field can only couple the states $|1,0,+\rangle$ and
$|0,0,-\rangle$ to second-order in the spin-orbit coupling for a
parabolic quantum dot;
 therefore, the EDSR technique cannot generate such transitions as efficiently as modulating $\omega_{X}$ when the spin-orbit coupling is weak.
\subsection{Increasing spin polarization in the lowest electronic
oribtal}  Throughout this work, relaxation has been neglected when
calculating the various spin transitions generated under parametric
modulation of the quantum dot's electrostatic potential.  In the
previous sections where only effective spin rotations within the
lowest electron state were considered, neglect of both $T_{1}$ and
$T_{2}$ relaxation seemed justified since the calculated Rabi
frequencies ($1-30$ MHz) were one to two orders of magnitude larger
than the measured\cite{Fujisawa02,Hanson03,Elzerman04,Amasha06}
$1/T_{1}$ values in quantum dots (we assume that $1/T_{2}$ is on the
same order\cite{Golovach04} as $1/T_{1}$). However, relaxation of
excited electronic states in quantum dots occurs on the nanosecond
time scale\cite{Petta04}, which is much faster than the calculated
Rabi frequencies (order of 10 MHz) associated with combined spin and
electronic excitation shown in Fig. \ref{fig:fig3};  therefore,
relaxation effects must be considered when examining such
transitions.

\begin{figure}%[b!]
\includegraphics*[height=10.7cm,width = 15.7cm]{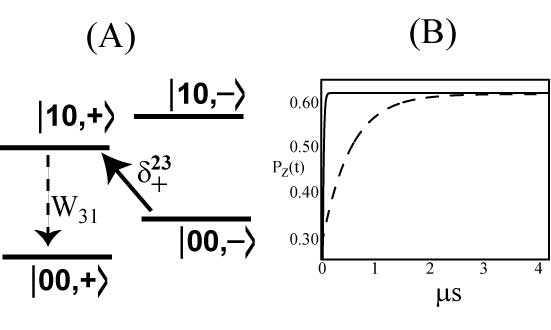}
\caption{Method for increasing the spin polarization of the lowest
electronic state, $|0,0\rangle$.  In (A), parametric modulations are
performed in order to coherently connect the states
$|2\rangle\equiv|0,0,-\rangle$ to the state
$|3\rangle\equiv|1,0,+\rangle$ (which is denoted by
$\delta_{\pm}^{23}$), thereby transferring population from state
$|2\rangle$ to state $|3\rangle$.  The population in state
$|3\rangle$ predominately relaxes to the state
$|1\rangle\equiv|0,0,+\rangle$ due to the electron directly coupling
to piezo-phonons, thereby increasing the spin polarization of the
ground electronic state,
$P_{Z}=\frac{p_{11}-p_{22}}{p_{11}+p_{22}}$.  (B) Integration of Eq.
(\ref{eq:eqmot}) gave $P_{Z}(t)$ for parametric modulations of
$\omega_{X}$ at the frequency $\omega_{r}=\omega_{X}-\omega_{Z}$
(solid curve) and at $\omega_{r}=(\omega_{X}-\omega_{Z})/2$ (dashed
curve). Along with the parameters used in Fig. \ref{fig:fig3}, the
following parameters were used in solving Eq. (\ref{eq:eqmot}):
 $W_{31}=3.74\times 10^8$ $s^{-1}$, $W_{32}=5\times 10^{-2}$
$s^{-1}$, $W_{21}=3.59\times 10^{3}$ $s^{-1}$, $T=2$ K,
$\Gamma_{23}=10 W_{31}$, and $|\delta_{\pm}^{23}|/(2\pi)\approx
48.1$ MHz (solid curve) and $|\delta_{\pm}^{23}|/(2\pi)\approx 8.77$
MHz (dashed curve).  In both cases,
$P_{Z}^{\text{steady-state}}\approx 0.62$ [Eq. (\ref{eq:pzss})],
although for larger $|\delta_{\pm}^{23}|$,
$P_{Z}^{\text{steady-state}}$ was reached on a faster time scale
[$0.15 \mu$s (solid curve) compared to $3\mu$s (dashed curve)].
 }\label{fig:figlast}
\end{figure}

The ability to coherently couple the states $|0,0,-\rangle$ and
$|1,0,+\rangle$, coupled with relaxation, can be used to help spin
polarize the electronic ground state of a quantum dot as depicted in
Fig. \ref{fig:figlast}(A). From Fig. \ref{fig:fig3}, parametric
modulation of the oscillator frequency can connect the state
$|0,0,-\rangle$ to the state $|1,0,+\rangle$ while leaving the state
$|0,0,+\rangle$ relatively uncoupled from any other state of the
quantum dot.  An electron in the excited state $|1,0,+\rangle$ can
quickly relax to the state $|0,0,+\rangle$ due to the direct
coupling of the electron to piezo-phonons\cite{Khaetskii01}, whereas
the relaxation of the state $|1,0,+\rangle$ to the state
$|0,0,-\rangle$, which requires an effective spin-phonon coupling,
occurs at a much smaller rate. This differential relaxation can be
used to generate spin polarization in the lowest electronic state in
a manner similar to optical pumping\cite{Happer72} and
Overhauser\cite{Slichter} techniques.

In the following work, the dynamics under parametric modulation of
the confining potential has been restricted to the following
subspace for simplicity: $\{|1\rangle\equiv
|0,0,+\rangle,\,|2\rangle\equiv |0,0,-\rangle,\,|3\rangle\equiv
|1,0,+\rangle\}$, and the calculations were performed using
different values of $\delta_{\pm}^{23}$ taken from the Fig.
\ref{fig:fig3}. The density matrix in this subspace can be written
as $\rho(t)=\sum_{i=1}^{3}p_{ii}(t)|i\rangle\langle
i|+p_{23}(t)|2\rangle\langle 3|+p_{32}(t)|3\rangle\langle 2|$ (where
only coherence between the states $|2\rangle$ and $|3\rangle$ have
been considered).  Defining $W_{ij}$ to be the transition rate from
state $i$ to state $j$, and $\Gamma_{23}$ to be the decoherence rate
for the coherence between states $|2\rangle$ and $|3\rangle$, the
various coefficients in $\rho(t)$ can be found by solving:
\begin{eqnarray}
\frac{\text{d}}{\text{d}t}\left(\begin{array}{c}p_{11}\\p_{22}\\p_{33}\\p_{23}\\p_{32}\end{array}\right)&=&\left(\begin{array}{ccccc}-(W_{12}+W_{13})&W_{21}&W_{31}&0&0\\
W_{12}&-(W_{21}+W_{23})&W_{32}&\text{i}\delta_{+}^{23}&-\text{i}\delta^{23}_{-}\\
W_{13}&W_{23}&-(W_{31}+W_{32})&-\text{i}\delta_{+}^{23}&\text{i}\delta_{-}^{23}\\
0&\text{i}\delta_{-}^{23}&-\text{i}\delta_{-}^{23}&-\Gamma_{23}&0\\
0&-\text{i}\delta^{23}_{+}&\text{i}\delta^{23}_{+}&0&-\Gamma_{23}\end{array}\right)\left(\begin{array}{c}p_{11}\\p_{22}\\p_{33}\\p_{23}\\p_{32}\end{array}\right)\nonumber\\
\label{eq:eqmot}
\end{eqnarray}
where the parametric modulation of the quantum dot is assumed not to
affect the values of $W_{ij}$ and $\Gamma_{23}$ in Eq.
(\ref{eq:eqmot}). In the absence of any term connecting the states
$|2\rangle$ and $|3\rangle$, i.e., $\delta_{\pm}^{23}=0$, the
various transition rates, $W_{ij}$, must satisfy the following
conditions in order to ensure that the equilibrium density matrix,
$\rho_{\text{eq}}=\sum_{i}p_{ii}^{\text{eq}}|i\rangle\langle i|$, is
a solution to Eq. (\ref{eq:eqmot}):
\begin{eqnarray}
W_{ij}&=&\frac{p_{jj}^{\text{eq}}}{p_{ii}^{\text{eq}}}W_{ji}
\end{eqnarray}
where $p_{jj}^{\text{eq}}$ is the equilibrium population for state
$j$, with $p_{11}^{\text{eq}}>p_{22}^{\text{eq}}>p_{33}^{\text{eq}}$
and $\sum_{i}p_{ii}^{\text{eq}}=1$.

In the presence of nonzero $\delta_{\pm}^{23}$, the effective spin
polarization of the ground electronic state,
$P_{Z}=\frac{p_{11}-p_{22}}{p_{11}+p_{22}}$, can increase from its
equilibrium value,
$P_{Z}^{\text{eq}}=\frac{p_{11}^{\text{eq}}-p_{22}^{\text{eq}}}{p_{11}^{\text{eq}}+p_{22}^{\text{eq}}}$.
Such an increase occurs since $W_{31}$ involves only orbital
relaxation, which is in general much faster than the transition rate
$W_{32}$, which requires both orbital relaxation and a spin flip.
Applying parametric modulations to the quantum dot can therefore
transfer population from state $|2\rangle$ to state $|3\rangle$,
which subsequently relaxes to state $|1\rangle$, thereby increasing
the population difference between states $|1\rangle$ and
$|2\rangle$. This is shown in Fig. \ref{fig:figlast}(A).

  The
steady-state spin polarization under such parametric oscillations
can be found by setting the left-hand side of Eq. (\ref{eq:eqmot})
to zero and solving for $p_{11}$ and $p_{22}$, which gives:
\begin{eqnarray}
P_{Z}^{\text{steady-state}}&=&\frac{W_{31}\left(W_{13}+W_{23}+2\frac{|\delta_{\pm}^{23}|^{2}}{\Gamma_{23}}\right)+(W_{21}-W_{12}-W_{13})\left(W_{31}+W_{32}+2\frac{|\delta_{\pm}^{23}|^{2}}{\Gamma_{23}}\right)}{W_{31}\left(W_{23}-W_{13}+2\frac{|\delta_{\pm}^{23}|^{2}}{\Gamma_{23}}\right)+(W_{21}+W_{12}+W_{13})\left(W_{31}+W_{32}+2\frac{|\delta_{\pm}^{23}|^{2}}{\Gamma_{23}}\right)}
\label{eq:pzss}
\end{eqnarray}
The transition rate, $W_{31}$, is mainly dominated by the coupling
of the electron to piezo-phonons.  From Eq. (4) of Ref.
\cite{Khaetskii00}, this rate is given as $W_{31}\approx 3.74\times
10^8$ $s^{-1}$ (for $\hbar\omega_{X}=0.25$ meV and
$\hbar\omega_{Y}=1$ meV);  using Eq. (7) of Ref. \cite{Khaetskii00}
as a rough estimate for $W_{32}$, one obtains $W_{32}\approx 5\times
10^{-2}$ $s^{-1}$, which satisfies $W_{31}\gg W_{32}$. Although no
theory has been attempted in this work to calculate $\Gamma_{23}$,
as long as $2\Gamma_{23}\geq W_{31}$, the diagonal elements,
$p_{ii}$, will be nonnegative (as seen from numerically integrating
Eq. (\ref{eq:eqmot})).  Although the value of $\Gamma_{23}$ does not
drastically affect $P_{Z}^{\text{steady-state}}$, it does affect the
time scale in which $P_{Z}^{\text{steady-state}}$ is reached.  In
the following we simply take, for illustrative purposes,
$\Gamma_{23}=10 W_{31}$. Finally, since the effective $T_{1}$ for
GaAs quantum dots is on the order of milliseconds to
microseconds\cite{Fujisawa02,Hanson03,Elzerman04,Amasha06}, a
conservative value of $T_{1}=100$ $\mu$s was chosen, which provided
the values of $W_{12}$ and $W_{21}$ from the condition that
$1/T_{1}=W_{12}+W_{21}=W_{21}(1+\exp(\hbar\omega_{Z}/(k_{B}T)))$
[where the populations were assumed to be given by the Boltzmann
distribution\cite{Beenakker91}].  For $T=2$ K and
$\hbar\omega_{Z}=0.1$ meV, $W_{21}\approx 3.59\times 10^3$ $s^{-1}$.
Using the parameters from Fig. \ref{fig:fig3}(A)
($|\delta_{\pm}^{23}|/(2\pi)=48.1$ MHz), $P_{Z}(t)$ was found by
integrating Eq. (\ref{eq:eqmot}) and is shown (solid curve) in Fig.
\ref{fig:figlast}(B), with $P_{Z}^{\text{stead-state}}\approx
0.62\approx 2.82 P_{Z}^{\text{eq}}$, which is reached within $0.15$
$\mu$s. No oscillations are present in $P_{Z}(t)$ since
$\Gamma_{23}\gg |\delta_{\pm}^{23}|$.

 For the case of modulating $\omega_{X}$ at a
frequency of $\omega_{r}\approx(\omega_{X}-\omega_{Z})/2$
 [with $|\delta_{2\omega,\pm}^{23}|/(2\pi)\approx 8.77$ MHz as shown in
Fig. \ref{fig:fig3}(B)], it took roughly twenty times as long
($\approx 3\mu$s) to reach $P_{Z}^{\text{steady-state}}\approx
0.62$, which is shown by the dashed curve in Fig.
\ref{fig:figlast}(B). It should be noted that
$P_{Z}^{\text{steady-state}}$ is approximately equal to the initial
equilibrium population difference between the states $|1\rangle$ and
$|3\rangle$,
$\frac{p_{11}^{\text{eq}}-p_{33}^{\text{eq}}}{p_{11}^{\text{eq}}+p_{33}^{\text{eq}}}\approx
0.62$.  Thus turning on the coupling between $|2\rangle$ and
$|3\rangle$ allows the initial ``orbital'' polarization between the
states $|1\rangle$ and $|3\rangle$ to be transferred into spin
polarization between the states $|1\rangle$ and $|2\rangle$.  This
process is similar to the Overhauser effect, where, by coupling the
electron and nuclear spins, the initial electron spin polarization
can be converted into nuclear polarization\cite{Slichter}.
Therefore, in order to increase $P_{Z}^{\text{steady-state}}$,
experiments should be performed at low temperatures, large magnetic
fields, and with increased lateral confinement, i.e., large
$\omega_{X}$.  However, increasing $\omega_{X}$ requires modulating
the quantum dot at a higher $\omega_{r}$ (which may be unfeasible
experimentally) in addition to the fact that $|\delta_{\pm}^{23}|$
decreases with increasing $\omega_{X}$, thereby increasing the time
it takes to reach $P_{Z}^{\text{steady-state}}$.

Finally, it should be noted that the above calculations assumed that
the quantum dot was isolated and  absolutely closed from the leads,
i.e., no electrons could enter or exit from the dot. However, if
electrons are able to tunnel in/out of the quantum dot, modulating
the electrostatic potential could be used to selectively ``kick''
out spin down electrons.  ``Re-zeroing'' the state of the quantum
dot to spin up could be used as an initialization step for a
possible quantum computation\cite{Friesen04,Elzerman04}. Such a
process would require the tunneling rate out of the dot for an
electron in state $|3\rangle$  to be much faster than $W_{31}$.
However, the evolution under parametric modulations of the confining
potential would have to be reconsidered, since in the case of a
non-isolated quantum dot,
 there would be a
finite probability during the parametric modulation that an electron
in the state $|1\rangle$ would get excited and kicked out of the dot
if the tunneling rate out of the dot is very large.
\section{Discussion and Conclusions}
In this work, a general formalism combining Floquet theory with
effective Hamiltonian theory was used to study the spin dynamics
under parametric modulations of a lateral quantum dot's
electrostatic potential in the presence of spin-orbit coupling. In
parabolic quantum dots, modulating the center of the quantum dot,
i.e., performing an EDSR experiment, was found to generate larger
Rabi frequencies than parametrically modulating the confining
frequency (in the absence of static electric fields), since the
latter is second-order in spin-orbit coupling.  However, in the
presence of a static electric field, both methods gave similar Rabi
frequencies for a parabolic quantum dot. For square dots, both EDSR
and modulating the width/length of the quantum dot generate Rabi
frequencies which are first-order in spin-orbit coupling.  The
modulation frequency must be chosen with precision on the order of
the Rabi frequency (on the order of tens of MHz) in order to
maximize the amplitude of the Rabi oscillations, thereby providing
better spectroscopic precision of the quantum dot's energy levels
relative to transport measurements, where thermal effects decrease
the spectral resolution.

Inclusion of the cubic Dresselhaus spin-orbit coupling didn't
dramatically alter the results obtained for parametric modulation in
a parabolic quantum dot.  However, the cubic Dresselhaus interaction
tends to decrease the contribution of the linear Dresselhaus
interaction to the Rabi frequency.  Furthermore, measurement of the
Rabi frequency under, for example, an EDSR experiment using
different orientations of the in-plane magnetic field could be used
to determine the ratio of the Rashba spin-orbit coupling constant to
the difference of the linear Dresselhaus spin-orbit coupling
constant with the product of the cubic Dresselhaus coupling constant
and the oscillator frequency ($\omega_{X}$ or $\omega_{Y}$).   For
example, using Eq. (\ref{eq:zz}) and including the cubic Dresselhaus
coupling from Eq. (\ref{eq:cub12ef}) gives:
\begin{eqnarray}
ZZ\left(\frac{\pi}{2}\right)&=&\frac{\beta-\frac{m^{*}\omega_{Y}\gamma}{2\hbar}}{\alpha}
\label{eq:zzc}
\end{eqnarray}
In order to separate $\gamma$ from $\beta$ in Eq. (\ref{eq:zzc}),
experiments could be repeated for different values of $\omega_{Y}$,
which would leave the linear Dresselhaus's contribution to
$ZZ(\pi/2)$ unchanged but would alter the cubic Dresselhaus's
contribution. Alternatively, it was shown that applying a
time-dependent electric field along the $\widehat{y}$ direction
coupled with parametric modulation of the confinement frequency
along the $\widehat{x}$ direction could be used to generate Rabi
frequencies which were proportional to $\gamma$ and were independent
of $\alpha$ and $\beta$ to first-order.  Measurement of the
resulting Rabi frequency would provide another independent measure
of the cubic Dresselhaus spin-orbit coupling constant.

Parametric modulations of the confining potential were also shown to
generate combined spin and orbital excitations in a parabolic
quantum dot with Rabi frequencies on the order of tens of MHz.
However, since electronic relaxation times are on the order of
nanoseconds, the effects of orbital relaxation had to be taken into
account.  A combination of coherent spin and orbital excitation with
orbital relaxation was shown to be able to increase the spin
polarization of the ground electronic state [Fig. \ref{fig:figlast}]
by transferring the initial ``electronic polarization'' between the
ground and excited electronic state into spin polarization of the
ground state.  For the parameters chosen in this work, this
corresponded to a three-fold increase in the spin polarization of
the ground electronic state. Larger spin polarizations could be
achieved by either using a more confined quantum dot (larger
$\omega_{X}$), larger magnetic fields, or by going to lower
temperatures.

Finally, the spin control developed in this work used only
parametric modulations of idealized electrostatic potentials
(parabolic and square-box), which should only be considered as a
model for spin control in lateral quantum dots.  For such methods to
be used in actual experimental quantum dots, more realistic
electrostatic potentials\cite{Stopa96} for a quantum dot, i.e.,
non-parabolic and non-square-box potentials, are required.
Furthermore, additional work is needed in order to better
characterize the actual time-dependent electrostatic potentials
generated by modulating the surface gate voltages of the quantum dot
along with designing optimal configurations of surface gates in
order to generate a desired transition. The results presented in
this work could furthermore be extended to many-electron quantum
dots, where the effects of electron-electron coupling on performing
spin excitations should be examined. Recent theoretical
work\cite{Trif07} has demonstrated that electronically controlled,
magnetic dipole-like couplings between spins in different quantum
dots can also be generated. These couplings, along with single spin
rotations, could be used to generate multiple-quantum coherences
between the electrons in different quantum dots by completely
electrostatic means.  In addition to being used to better
characterize the electron-environment coupling of such highly
correlated states, generating such highly correlated states could be
used as an intermediary step during a quantum computation.

\acknowledgments{We would like to thank Prof. Eric Heller and Prof.
Yung-Ya Lin for their support. This work was supported by NSF
(CHE-0349362 and CHE-0116853).

%\bibliographystyle{prstya}
%\bibliography{hellerbib}

\appendix
\section{The Effective Hamiltonian}
If for a given Hamiltonian (written for convenience in Floquet
space), $\widehat{H}_{F}=\widehat{H}_{F}^{0}+\widehat{V}_{F}$, there
exists a subspace of interest, $Q$, which is weakly coupled by
$\widehat{V}_{F}$ to the rest of the Floquet space, $U$, then the
dynamics within $Q$ can be separated from $U$ by constructing an
effective Hamiltonian from $\widehat{H}_{F}$,
$\widehat{H}_{F}^{\text{EFF}}$, which is block-diagonal in the $Q$
and $U$ subspaces\cite{Shavitt80,Tannoudji}. Defining
$\widehat{P}_{Q}$ to be the projection operator onto the subspace
$Q$, $\widehat{P}_{Q}=\sum_{|\alpha,m_{F}\rangle\in
Q}|\alpha,m_{F}\rangle\langle\alpha,m_{F}|$, and $\widehat{P}_{U}$
to be the complementary projection operator onto the subspace $U$,
$\widehat{P}_{U}=\widehat{1}_{F}-\widehat{P}_{Q}$,
$\widehat{H}_{F}^{\text{EFF}}$ can be determined by constructing a
unitary transformation, $\exp(\widehat{S}_{F})$, such that
$\exp\left(\widehat{S}_{F}\right)\widehat{H}_{F}\exp\left(-\widehat{S}_{F}\right)=\widehat{H}^{\text{EFF}}_{F}$,
where in order to ensure that $\widehat{H}_{F}^{\text{EFF}}$ is
block diagonal in both the $Q$ and $U$ subspaces,
$\widehat{H}_{F}^{\text{EFF}}$ must satisfy the following:
\begin{eqnarray}
\widehat{P}_{Q}\widehat{H}^{\text{EFF}}_{F}\widehat{P}_{U}=\widehat{P}_{U}\widehat{H}^{\text{EFF}}_{F}\widehat{P}_{Q}=\widehat{0}
\label{eq:heffdeff}
\end{eqnarray}
Using Eq. (\ref{eq:heffdeff}), a perturbation expansion for
$\widehat{S}_{F}$ in powers of $\widehat{V}_{F}$,
$\widehat{S}_{F}=\sum_{m=1}^{\infty}\widehat{S}_{F}^{(m)}$, can be
constructed.
 Separating $\widehat{V}_{F}$ into its diagonal and off-diagonal
 components, $\widehat{V}_{F}=\widehat{V}^{S}_{F}+\widehat{V}^{D}_{F}$,
 with $\widehat{V}^{S}_{F}=\widehat{P}_{Q}\widehat{V}_{F}\widehat{P}_{U}+\widehat{P}_{U}\widehat{V}_{F}\widehat{P}_{Q}$
and
$\widehat{V}^{D}_{F}=\widehat{P}_{U}\widehat{V}_{F}\widehat{P}_{U}+\widehat{P}_{Q}\widehat{V}_{F}\widehat{P}_{Q}$,
$\widehat{S}_{F}$ can be written up to $m=3$ as:
\begin{eqnarray}
 \langle
\alpha,m_{F}|\widehat{S}^{(1)}_{F}|\beta,n_{F}\rangle&=&\frac{\langle
\alpha,m_{F}|\widehat{V}^{S}_{F}|\beta,n_{F}\rangle}{E_{\alpha}-E_{\beta}+\hbar(m-n)\omega_{r}}\nonumber\\
\langle\alpha,m_{F}|\widehat{S}^{(2)}_{F}|\beta,n_{F}\rangle&=&\frac{\langle\alpha,m_{F}|[\widehat{S}^{(1)}_{F},\widehat{V}^{D}_{F}]|\beta,n_{F}\rangle}{E_{\alpha}-E_{\beta}+\hbar(m-n)\omega_{r}}\nonumber\\
\langle\alpha,m_{F}|\widehat{S}^{(3)}_{F}|\beta,n_{F}\rangle&=&\frac{\langle\alpha,m_{F}|[\widehat{S}^{(2)}_{F},\widehat{V}^{D}_{F}]|\beta,n_{F}\rangle}{E_{\alpha}-E_{\beta}+\hbar(m-n)\omega_{r}}+\frac{\langle
\alpha,m_{F}|[\widehat{S}^{(1)}_{F},[\widehat{S}^{(1)}_{F},V^{S}_{F}]]|\beta,n_{F}\rangle}{3\left(E_{\alpha}-E_{\beta}+\hbar(m-n)\omega_{r}\right)}
\label{eq:sexp}
\end{eqnarray}
Using the above values of $\widehat{S}_{F}$ in Eq. (\ref{eq:sexp}),
the effective Hamiltonian in the subspace $Q$,
$\widehat{P}_{Q}\widehat{H}_{F}^{\text{EFF}}\widehat{P}_{Q}=\sum_{m=0}^{\infty}\widehat{H}^{\text{EFF}(m)}_{F}$,
is given (up to order $(\widehat{V}_{F})^4$) as:
\begin{eqnarray}
\langle\alpha_{1},n_{F}|\widehat{H}^{\text{EFF}(0)}_{F}|\alpha_{2},m_{F}\rangle&=&\langle\alpha_{1},n_{F}|\widehat{H}^{0}_{F}|\alpha_{2},m_{F}\rangle\nonumber\\
\langle\alpha_{1},n_{F}|\widehat{H}^{\text{EFF}(1)}_{F}|\alpha_{2},m_{F}\rangle&=&\langle\alpha_{1},n_{F}|\widehat{V}^{D}_{F}|\alpha_{2},m_{F}\rangle\nonumber\\
\langle\alpha_{1},n_{F}|\widehat{H}^{\text{EFF}(2)}_{F}|\alpha_{2},m_{F}\rangle&=&\frac{1}{2}\langle\alpha_{1},n_{F}|[\widehat{S}^{(1)}_{F},\widehat{V}^{S}_{F}]|\alpha_{2},m_{F}\rangle\nonumber\\
\langle\alpha_{1},n_{F}|\widehat{H}^{\text{EFF}(3)}_{F}|\alpha_{2},m_{F}\rangle&=&\frac{1}{2}\langle\alpha_{1},n_{F}|[\widehat{S}^{(2)}_{F},\widehat{V}^{S}_{F}]|\alpha_{2},m_{F}\rangle\nonumber\\
\langle\alpha_{1},n_{F}|\widehat{H}^{\text{EFF}(4)}_{F}|\alpha_{2},m_{F}\rangle&=&\frac{1}{2}\langle\alpha_{1},n_{F}|[\widehat{S}^{(3)}_{F},\widehat{V}^{S}_{F}]|\alpha_{2},m_{F}\rangle-\frac{1}{24}\langle\alpha_{1},n_{F}|[\widehat{S}^{(1)}_{F},[\widehat{S}^{(1)}_{F},[\widehat{S}^{(1)}_{F},\widehat{V}^{S}_{F}]]]|\alpha_{2},m_{F}\rangle\nonumber\\
\end{eqnarray}
\end{document}